\documentclass[a4paper,fleqn,usenatbib]{mnras}

\usepackage{newtxtext,newtxmath}

\usepackage[T1]{fontenc}
\usepackage{ae,aecompl}

\usepackage{graphicx}	
\usepackage{amsmath}	
\usepackage{amssymb}	
\usepackage{setspace}
\usepackage{multicol}
\usepackage{ragged2e}
\usepackage{xcolor}

\title[Modeling the light curve of `Oumuamua]{Modeling the light curve of `Oumuamua: evidence for torque and disc-like shape}

\author[S. Mashchenko]{
Sergey Mashchenko\thanks{E-mail: syam@physics.mcmaster.ca}
\\
Department of Physics and Astronomy, McMaster University, 1280 Main Street West, Hamilton ON L8S 4M1, Canada \\
}

\date{Accepted 2019 August 22. Received 2019 August 15; in original form 2019 June 9}

\pubyear{2019}
\volume{489}
\pagerange{3003-3021}

\begin{document}
\label{firstpage}
\maketitle

\begin{abstract}

We present the first attempt to fit the light curve of the interstellar visitor `Oumuamua using a physical model which includes optional torque.
We consider both conventional (Lommel-Seeliger triaxial ellipsoid) and alternative
("black-and-white ball", "solar sail") brightness models. With all the brightness models, some torque is required to explain the
timings of the most conspicuous features -- deep minima -- of the asteroid's light curve. Our best-fitting models are a thin disc
(aspect ratio 1:6) and a thin cigar (aspect ratio 1:8) which are very close to being axially symmetric. 
Both models are tumbling and require some torque which has the same amplitude in relation
to `Oumuamua's linear non-gravitational acceleration as in Solar System comets which dynamics is affected by outgassing. 
Assuming random orientation of the angular momentum vector, we compute probabilities for our best-fitting models. We show
that cigar-shaped models suffer from a fine-tuning problem and have only 16~per cent probability to produce light curve minima as deep
as the ones present in `Oumuamua's light curve. 
Disc-shaped models, on the other hand, are very likely (at 91~per cent) to produce
minima of the required depth. From our analysis, the most likely model for `Oumuamua is a thin disc (slab) experiencing moderate torque
from outgassing.

\end{abstract}




\section{Introduction}

1I/2017 `Oumuamua is the first and only known interstellar minor body to pass through the Solar System. It was detected by Pan-STARRS1 survey
on October 19, 2017, and by October 22 was determined to have by far the largest known hyperbolic eccentricity of 1.2 \citep{Meech2017}.
Unfortunately it was discovered when it was already on its way out of the Solar System, after passing its perihelion ($0.25$~au from the Sun) on September 9, 
and having a close approach (0.16~au) to Earth on October 14. This severely limited the number of observations which could be acquired.
Most of observations of `Oumuamua were done during the 5-day interval between October 25
and October 30 \citep{Fraser2018,Drahus2018}, with a few more observations over the next two months until its final sighting on January 2, 2018 \citep{Micheli2018}.
The observations were primarily done in visible light, though two very interesting non-detections in other wavelengths were also reported:
in infrared by {\it Spitzer} \citep{Trilling2018} and in radio by SETI \citep{Harp2019}.

The second unique feature of `Oumuamua (in addition to its interstellar nature) was its extreme brightness variability. At $2.6\pm 0.2$~mag,
the amplitude of the brightness changes was larger than for any Solar System minor body, suggesting an extreme geometry \citep{Drahus2018}. The early suggestion
was that `Oumuamua has a very prolate (cigar-like) shape \citep{Meech2017} which was later accepted by most of the literature on this object,
though very oblate (disc-like) shapes would work as well \citep{Belton2018}. The idea that `Oumuamua's large brightness variations are not geometric in nature, but are primarily
driven by large albedo variations across the asteroid's surface, was briefly entertained and dismissed as unlikely due to the absence of any sign of volatiles \citep{Meech2017}.

`Oumuamua's light curve was also unusual for another reason. While early papers based on limited data suggested that the asteroid is a simple rotator 
with the rotation period between 7.3 and 8.1~h \citep{Meech2017,Jewitt2017,Bolin2018}, later papers which analysed more complete datasets 
concluded (by analysing the periodograms for the light curve) that the asteroid is in an excited (Non Principal Axis, or tumbling) rotational state \citep{Belton2018,Drahus2018,Fraser2018}. 
It is important to note that periodograms
of noisy and patchy data of a limited size can produce fake dominant frequencies \citep{Samarasinha2015}. Also, such an analysis assumes there is no torque.
Dominant frequencies found in periodograms should be treated as suggestive only, and should be ideally verified (confirmed or disproved) by means of physical modeling
of the light curve \citep{Pravec2005}. 

The third unique feature of `Oumuamua is its non-gravitational acceleration, discovered by \citet{Micheli2018}, combined with the lack of any signs of outgassing 
\citep{Drahus2018,Trilling2018,Sekanina2019}. In Solar System comets, non-gravitational acceleration is usually associated with active outgassing.  
This conundrum spurred some non-orthodox explanations, such as the solar sail idea of \citet{Bialy2018}. 

As evidenced by Solar System comets and expected on theoretical grounds,
the same agent (e.g. outgassing), which drives linear non-gravitational acceleration of a minor body, should also produce torque, which amplitude 
should correlate with the amplitude of the linear acceleration \citep{Rafikov2018a}. \citet{Seligman2019} showed that their physical model of `Oumuamua, where
the outgassing point tracks the subsolar spot on the asteroid's surface, can reproduce the magnitude of the non-gravitational linear acceleration 
and some features of the light curve. (We have to emphasize that the authors did not carry out computationally expensive fitting of the
observed light curve.) On the other hand, recently \citet{Rafikov2018b}, based on the frequency analysis of the light curve by \citet{Belton2018}, claimed that `Oumuamua
should have experienced negligible torque. We critically assess this claim in our paper.

Despite significant research efforts, the nature of `Oumuamua remains a puzzle. 
Perhaps it is a comet, which would be in line with the theoretical expectations predicting $200\dots 10^4$ times more icy objects than
rocky objects among interstellar minor bodies \citep{Meech2017}, and because it exhibited strong non-gravitational acceleration. Or perhaps it is an asteroid,
judging from the non-detection of any signs of outgassing, but then the non-gravitation acceleration remains unexplained. Or it could be something else, e.g. a solar sail.
Clearly more efforts are needed to bring some clarity to this issue.

This paper represents the first attempt to fit a physical model to the observed light curve of `Oumuamua, with all free model parameters recovered by means of multi-dimensional optimization.
The paper is organized as follows. Section~\ref{sec:model} describes the two main components of the model: the kinematic part (spin evolution of a tumbling asteroid with optional
constant torque), and the brightness model part (can be either a geometric one -- Lommel-Seeliger triaxial ellipsoid, or a variable albedo one -- "black-and-white ball").
Section~\ref{sec:code} presents our GPU-based numerical code, describes the numerical setup, and details code validation tests. Section~\ref{sec:oumuamua} describes the observational
data used for modeling, and presents the results of fitting `Oumuamua's light curve with our physical model (with and without torque). 
The paper ends with Discussion (\autoref{sec:discussion}) and Conclusions and Future Work (\autoref{sec:conclusions}).

\section{Model}
\label{sec:model}

\subsection{Overview}

Our model consists of two major components: kinematic model, and brightness model. 

In terms of kinematics, `Oumuamua is assumed to be a rigid body
with an arbitrary shape and arbitrary density distribution, which rotates in a Non Principal Axis (NPA) mode; in other words, it is tumbling.
As the simplest non-inertial extension, the model can optionally account for arbitrary torque which is fixed in time and 
space (in the asteroidal comoving coordinate system). Physically, this might correspond to semi-steady outgassing from a specific point on the
asteroid's surface. We describe the equations of motion in \autoref{sec:eq_motions}, the initial conditions in \autoref{sec:ICs}, and the
model's coordinate transformations in \autoref{sec:coords}.

Our main brightness model (described in \autoref{sec:br_model}) assumes that the asteroid is a triaxial ellipsoid with uniform albedo 
surface with Lommel-Seeliger (LS) light scattering properties.
The ellipsoid can be either self-consistent (with the semi-axes lengths taken from the kinematic part of the model),
or relaxed (with the semi-axes lengths not linked to the kinematic model). 
Relaxing the brightness ellipsoid parameters can help to account for potential deviations of the asteroid's properties (e.g. shape) from the model 
assumptions \citep{Pravec2005}.

We also explore the simplest non-geometric
explanation for the large brightness variations of `Oumuamua: a spherical body with the two hemispheres having different albedo values, which is
oriented arbitrarily relative to the diagonal components of the inertia tensor. This model is described in \autoref{sec:BW}.

\subsection{Equations of motion}
\label{sec:eq_motions}

We adopt a comoving right-handed Cartesian coordinate system with the three principal axes -- $\mathbfit{b}$, $\mathbfit{c}$, and $\mathbfit{a}$ -- coinciding with the
three diagonal components of the asteroid's inertia tensor, $I_b$, $I_c$, and $I_a$, respectively. The axes are chosen
in a way that the following inequalities are always true: $I_a \lid I_b \lid I_c$. (Our axes $\mathbfit{b}$, $\mathbfit{c}$, and $\mathbfit{a}$
are equivalent to the axes $\mathbfit{i}$, $\mathbfit{s}$, and $\mathbfit{l}$ of \citealt{Samarasinha1991}.) If the asteroid's shape can be described as a 
triaxial ellipsoid, the corresponding semi-axes of the ellipsoid would 
follow the $a \gid b \gid c$ relations.

We adopt the units where $a=1$ and $I_a=1$; the time unit is a day. In these units, the three diagonal components of the inertia tensor of a triaxial ellipsoid are

\begin{equation}
    I_b = \frac{1 + c^2}{b^2 + c^2},\label{eq:Iabc}\\
    I_c = \frac{1 + b^2}{b^2 + c^2},\\
    I_a = 1.
\end{equation}

In the comoving coordinate system, Euler's equations for rigid body rotation can be written as 

\begin{equation}
    \begin{array}{lcl}
    I_b \dot{\Omega}_b + \left(1-I_c\right)\Omega_c\Omega_a  &=& K_b,\\
    I_c \dot{\Omega}_c + \left(I_b-1\right)\Omega_a\Omega_b  &=& K_c,\label{eq:euler3}\\
    \dot{\Omega}_a + \left(I_c-I_b\right)\Omega_b\Omega_c  &=& K_a\\
    \end{array}
\end{equation}

\noindent  \citep[][p. 115]{Landau1976Mechanics}. Here ${\Omega}_b$, ${\Omega}_c$, ${\Omega}_a$ and $K_b$, $K_c$, $K_a$ are the components  of the 
angular velocity vector and the torque pseudo vector, respectively, in the comoving (asteroidal) coordinate system. The angular velocity vector components can be expressed in terms
of the three Euler angles (nutation angle $\theta$, precession angle $\varphi$, and rotation angle $\psi$) and their derivatives:

\begin{equation}
    \begin{array}{lcl}
    \Omega_b &=& \dot{\varphi} \sin\theta\sin\psi + \dot{\theta}\cos\psi,\\
    \Omega_c &=& \dot{\varphi} \sin\theta\cos\psi - \dot{\theta}\sin\psi,\label{eq:omega}\\
    \Omega_a &=& \dot{\varphi} \cos\theta + \dot{\psi}
    \end{array}
\end{equation}

\noindent  \citep[][p. 111]{Landau1976Mechanics}.

Equations~(\ref{eq:euler3}) and (\ref{eq:omega}) can be rewritten to form a system of six ordinary differential equations (ODEs) for the three $\Omega$ components and the three Euler angles:

\begin{equation}
    \left\{
    \begin{array}{lcl}
    \dot{\Omega}_b &=& \Omega_c\Omega_a \left(I_c-1\right)/I_b + T_b,\\
    \dot{\Omega}_c &=& \Omega_a\Omega_b \left(1-I_b\right)/I_c + T_c,\\
    \dot{\Omega}_a &=& \Omega_b\Omega_c\left(I_b-I_c\right) + T_a,\\
    \dot{\varphi}  &=& \left(\Omega_b\sin\psi + \Omega_c\cos\psi\right)/\sin\theta,\label{eq:euler2}\\
    \dot{\theta} &=& \Omega_b\cos\psi - \Omega_c\sin\psi,\\
    \dot{\psi} &=& \Omega_a - \dot{\varphi}\cos\theta.
    \end{array}
    \right.
\end{equation}

\noindent Here $T_b=K_b/I_b$, $T_c=K_c/I_c$, and $T_a=K_a$ are the components of the torque vector normalized by the corresponding diagonal components of the inertia tensor.

The system~(\ref{eq:euler2}) is the one we need to integrate numerically to describe the rotation of a rigid body in the presence of torque. If torque is zero,
there is a trick allowing one to bypass the Euler equations, and reduce the problem to only three ODEs, for the three Euler angles \citep{Kaasalainen2001}.
Specifically, in the torque-free regime the angular momentum vector $\mathbfit{L}$ of a rotating rigid body is fixed in all inertial coordinate systems 
(angular momentum conservation), which lets us write the following equations:

\begin{equation}
    \begin{array}{lcl}
    \Omega_b &=& L I_b^{-1} \sin \theta \sin \psi,\\
    \Omega_c &=& L I_c^{-1} \sin \theta \cos \psi,\label{eq:L}\\
    \Omega_a &=& L \cos \theta
    \end{array}
\end{equation}

\noindent  \citep[][p. 119]{Landau1976Mechanics}. As shown by \citet{Kaasalainen2001}, combining equations~(\ref{eq:omega}) and (\ref{eq:L}) results in the
following system of three ODEs, for the three Euler angles:

\begin{equation}
    \left\{
    \begin{array}{lcl}
    \dot{\varphi}   &=& L (I_+ - I_- \cos 2\psi),\\
    \dot{\theta} &=& L I_- \sin \theta \sin 2\psi,\label{eq:euler1}\\
    \dot{\psi}   &=& \cos \theta (L - \dot{\varphi}).
    \end{array}
    \right.
\end{equation}

\noindent Here $I_- = \frac12 \left(I_b^{-1}-I_c^{-1}\right)$ and $I_+ = \frac12 \left(I_b^{-1}+I_c^{-1}\right)$.

\citet{Kaasalainen2001} set the principal axes of the comoving coordinate system differently for Short Axis Mode (SAM) and Long Axis Mode (LAM) rotators, 
which allowed them to simplify many model equations, with only one form of an equation for both SAM and LAM cases. In our testing,
this worked well for mildly flattened objects. Unfortunately, for the shortest-to-longest ellipsoid axes ratios ($c/a$) smaller than $\approx 0.2$ 
we observed the ODEs integration errors to quickly become significant (necessitating much smaller time steps, which would make simulations much longer). 
At some point (around $c/a\la 0.15$), the ODEs integration completely breaks down due to some numerical instability. No such issues were observed when we used the same
comoving axes assignment (with $c$ and $a$ always corresponding to the smallest and largest ellipsoid's semi-axes, respectively), for both SAM and LAM objects, 
as in \citet{Samarasinha1991}. Using this latter approach allowed us to use a fairly large integration time step
without noticeably affecting the accuracy of integration. Crucially, this also allowed us to fully explore  the range of $c/a$ ratios
needed to explain `Oumuamua's extreme brightness variations.

In our model, we solve equations~(\ref{eq:euler1}) for torque-free runs, and equations~(\ref{eq:euler2}) for runs with torque (which is assumed to be constant in the 
comoving coordinate system). It is important to emphasize that the equations are applicable to any rigid body (not just triaxial ellipsoids), described by the three diagonal
components of its inertia tensor --  $I_b$, $I_c$, and $I_a$.

\subsection{Initial conditions}
\label{sec:ICs}

To solve either equations~(\ref{eq:euler1}) or equations~(\ref{eq:euler2}), one has to set the initial values of the independent variables. In the adopted comoving coordinate system, $bca$, 
precession angle $\varphi$ initially can have any value: $\varphi_0 \in [0,2\pi]$. 

Angular momentum vector modulus $L$ can have any positive value initially, $L\in[0,\infty[$. Total allowed range for the model parameter $E' \equiv 2E/L^2$ is $I_c^{-1} \ldots 1$. 
(Here $E$ is the rotational kinetic energy of the body.) In the short and long axis modes (SAM and LAM), the corresponding sub-ranges are $I_c^{-1} \ldots I_b^{-1}$ 
and $I_b^{-1} \ldots 1$, respectively \citep[equations~A30 and A54]{Samarasinha1991}. Parameters $L$ and $E'$ are fixed in torque-free simulations, but change with time in runs with non-zero torque. In the latter case, only the initial values of the two parameters need to be provided; the ODEs (\autoref{eq:euler2}) do not explicitly use them.

Rotation angle $\psi$ can have any value ($\psi_0 \in [0,2\pi]$) in LAM, but is constrained to the following range in SAM \citep[equations~A63 and A64]{Samarasinha1991}:

\begin{equation}
 \setstretch{3}
   \begin{array}{lcr}
    \psi_{\rm min} &=& -\arctan \left[ \sqrt{\dfrac{I_b (I_c-1/E')}{I_c (1/E'-I_b)}} \right],\\
    \psi_{\rm max} &=& \arctan \left[ \sqrt{\dfrac{I_b (I_c-1/E')}{I_c (1/E'-I_b)}} \right]. \label{eq:psi}
    \end{array}
\end{equation}

Combining the expressions for the components of the angular momentum vector,

\begin{equation}
    \begin{array}{lcl}
    L_b &=& L \sin \theta \sin \psi,\\
    L_c &=& L \sin \theta \cos \psi,\label{eq:L3}\\
    L_a &=& L \cos \theta
    \end{array}
\end{equation}

\noindent  \citep[][p. 119]{Landau1976Mechanics} with the kinetic energy equation,

\begin{equation}
    2 E = \frac{L_b^2}{I_b} + \frac{L_c^2}{I_c} + L_a^2 \label{eq:E}
\end{equation}

\noindent  \citep[][p. 116]{Landau1976Mechanics} allows us to write the expression for the initial value of the nutation angle:

\begin{equation}
    \theta_0 = \arcsin \left[ \sqrt{\frac{E'-1}{\sin^2\psi_0 (I_b^{-1}-I_c^{-1})+I_c^{-1}-1}} \right].\label{eq:theta0}
\end{equation}

\noindent As one can see, $\theta_0$ is not a free parameter (unlike $\varphi_0$ and $\psi_0$): it is fully determined by other model parameters ($E'$, $\psi_0$, $b$ and $c$).
Nutation angle can vary between $0$ and $\pi$.

Once we set the values for the parameters $L$, $\theta_0$, and $\psi_0$, the initial values of the three components of the angular velocity vector $\Omega$ for simulations with non-zero torque can be computed using equations~(\ref{eq:L}).

When modeling light curves of asteroids, it is common to use rotation period $P_\psi$ and precession period $P_\varphi$ (which can often be deduced or guessed from 
observations) in place of the physical parameters $E$ (or $E'$ in our case) and $L$. This necessitates the conversion 
($P_\psi$, $P_\varphi$) $\rightarrow$ ($E'$, $L$) for each tested model which
is computationally expensive and can dramatically slow down the simulations. We use a compromise approach in our model, 
which can take either $L$ or $P_\psi$ as a free parameter.
We keep $E'$ (which is relatively well constrained) as another free parameter. When $P_\psi$ is a free parameter, we derive $L$ from $P_\psi$ and $E'$ using
the following efficient computational routine.

First we compute the parameter $k^2$ \citep[equations~A32 and A56]{Samarasinha1991}:

\begin{equation}
\setstretch{3}
k^2 = \left\{
    \begin{array}{lcl}
    \dfrac{(I_c-I_b)(1/E'-1)}{(I_b-1)(I_c-1/E')}  {\rm\quad(LAM)},\\
    \dfrac{(I_b-1)(I_c-1/E')}{(I_c-I_b)(1/E'-1)}  {\rm\quad(SAM)}. \label{eq:k2}
    \end{array}
    \right.
\end{equation}

\noindent Next we compute the elliptic integral,

\begin{equation}
    K_e = \int_0^{\pi/2} \frac{du}{\sqrt{1-k^2 \sin^2 u}} \label{eq:K}
\end{equation}

\noindent \citep[][p. 118]{Landau1976Mechanics}, using very efficient AGM (arithmetic-geometric mean) iterative method\footnote{\url{https://en.wikipedia.org/wiki/Arithmetic-geometric_mean}}:

\begin{verbatim}
    for (int i=0; i<N; i++)
    {
        a1 = (a+g)/2;
        g1 = sqrt(a*g);
        a = a1;  g = g1;
    }
\end{verbatim}

\noindent If initial values of the variables $a$ and $g$ are set to $1$ and $\sqrt{1-k^2}$, respectively, the above iterative loop quickly converges, with $\pi/(a+g)\rightarrow K_e$
as $N\rightarrow\infty$. In our testing, after only 5 iterations the error in $K_e$ is smaller than $10^{-10}$, for $k^2=0\ldots0.9999998$.

Now we can derive $L$ from $P_\psi$ as follows:

\begin{equation}
\setstretch{3}
L = \left\{
    \begin{array}{lcl}
     \dfrac{4K_e}{P_\psi} \sqrt{\dfrac{I_b I_c}{E'(I_b-1)(I_c-E'^{-1})}}  {\rm\quad(LAM)},\\
      \dfrac{4K_e}{P_\psi} \sqrt{\dfrac{I_b I_c}{E'(I_c-I_b)(E'^{-1}-1)}}  {\rm\quad(SAM)} \label{eq:L_Ppsi}
   \end{array}
    \right.
\end{equation}

\noindent \citep[equations~A45 and A71]{Samarasinha1991}.

\subsection{Coordinate transformations}
\label{sec:coords}

Our model utilizes three different right-handed Cartesian coordinate systems. The starting point is the inertial Solar System Barycentre (SSB) coordinate system,
$xyz$. We used online NASA's tool HORIZONS\footnote{\url{https://ssd.jpl.nasa.gov/horizons.cgi}} \citep{Giorgini1996} to generate SSB coordinates for the centres of the Sun, Earth, and `Oumuamua for all the data points 
in `Oumuamua's light curve.

\begin{figure}
    \includegraphics[width=\columnwidth]{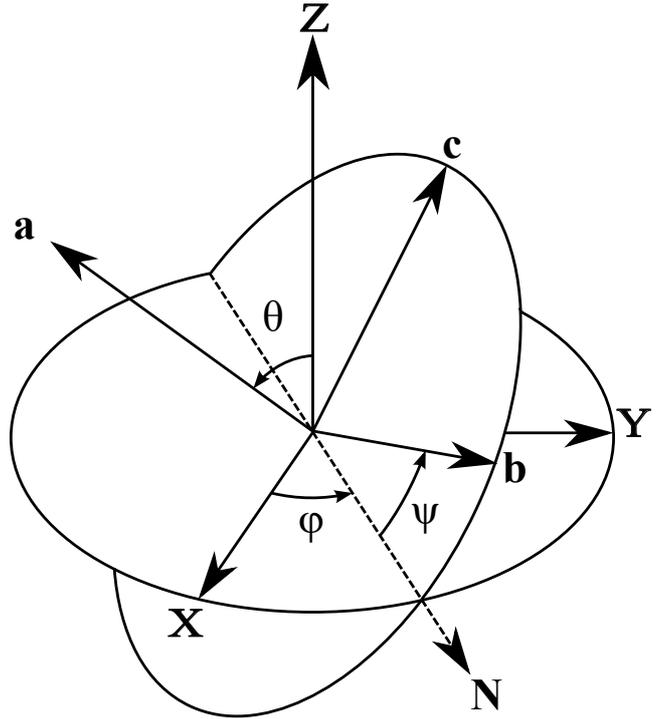}
    \caption{Transformation from the inertial coordinate system $XYZ$ ($Z$ being the initial orientation of the
    angular momentum vector $\mathbfit{L}$) to the comoving coordinate system $bca$ using Euler angles $\theta$, $\varphi$, and $\psi$.}
    \label{fig:Eulerangles}
\end{figure}

The second coordinate system, $XYZ$, is also inertial. The axis $\mathbfit{Z}$ coincides with the angular momentum vector $\mathbfit{L}$. 
(For runs with non-zero torque, $\mathbfit{Z}$ coincides with
the angular momentum vector $\mathbfit{L}$ at the initial moment of time.) The axis $\mathbfit{X}$ is arbitrarily chosen to coincide with the vector $\mathbfit{y} \boldsymbol\times \mathbfit{Z}$. 
The axis $\mathbfit{Y}$ complements the other two axes to form a right-handed coordinate system: $\mathbfit{Y}=\mathbfit{Z}\boldsymbol\times \mathbfit{X}$. 
The three axes of the $XYZ$ coordinate system can be described as unit vectors in the SSB ($xyz$) coordinate system as follows:

\begin{equation}
\setstretch{2}
    \begin{array}{lcl}
    Z_{x,y,z} &=& \left\{\sin\theta_L \cos\varphi_L,\> \sin\theta_L \sin\varphi_L,\> \cos\theta_L\right\},\\
    X_{x,y,z} &=& \left\{Z_z/\sqrt{Z_z^2+Z_x^2},\> 0,\> -Z_x/\sqrt{Z_z^2+Z_x^2}\right\} \label{eq:XYZ},\\
    Y_{x,y,z} &=& \left\{Z_y X_z,\> Z_z X_x-Z_x X_z,\> -Z_y X_x\right\}.
    \end{array}
\end{equation}

\noindent Here free model parameters $\theta_L$ and $\varphi_L$ are polar and azimuthal angles, respectively, describing the (initial) orientation of the angular momentum 
vector $\mathbfit{L}$ in the
SSB coordinate system.

Finally, the comoving (asteroidal) coordinate system, $bca$, has its three axes coinciding with the intermediate, largest, and smallest
diagonal components of the asteroidal inertia tensor. This coordinate system is derived by rotating the $XYZ$ coordinate system using three Euler angles,
$\theta$, $\varphi$, and $\psi$ (\autoref{fig:Eulerangles}), which are derived by solving the equations of motion (\autoref{sec:eq_motions}).
If the mass distribution of the asteroid can be well approximated as a homogeneous triaxial 
ellipsoid, the three axes correspond to the intermediate ($b$), smallest ($c$), and largest ($a$) semi-axes of the ellipsoid. The SSB components of the three
axes, $\mathbfit{b}$, $\mathbfit{c}$, and $\mathbfit{a}$, can be computed via a sequence of geometric transformations as follows.

Components of the unit node vector $\mathbfit{N}$, derived by rotating vector $\mathbfit{X}$ towards vector $\mathbfit{Y}$ by the Euler angle $\varphi$ (see~\autoref{fig:Eulerangles}),
with $\mathbfit{Z}$ being the rotation axis,
in the SSB coordinate system are given by

\begin{equation}
    \begin{array}{lcr}
    N_{x,y,z} &=& \left\{X_x \cos\varphi+Y_x \sin\varphi,\> Y_y \sin\varphi,\right.\\
    &&  \left. X_z \cos\varphi+Y_z \sin\varphi\right\}.
    \label{eq:N}
    \end{array}
\end{equation}

\noindent Using another auxiliary unit vector, $\mathbfit{p}=\mathbfit{N} \boldsymbol\times \mathbfit{Z}$, 
            
\begin{equation}
    p_{x,y,z} = \left\{N_y Z_z-N_z Z_y,\>  N_z Z_x - N_x Z_z,\> N_x Z_y - N_y Z_x\right\}, \label{eq:p}
\end{equation}

\noindent allows us to derive the axis $\mathbfit{a}$ (a unit vector) components in the SSB coordinate system by rotating  $\mathbfit{Z}$ by the Euler angle 
$\theta$ towards $\mathbfit{p}$, with the node vector $\mathbfit{N}$ being the rotation vector:

\begin{equation}
    \begin{array}{lcr}
    a_{x,y,z} &=& \left\{Z_x\cos\theta + p_x\sin\theta,\>  Z_y\cos\theta + p_y\sin\theta,\right.\\
    &&  \left. Z_z\cos\theta + p_z\sin\theta\right\}.
    \label{eq:a}
    \end{array}
\end{equation}

\noindent Using yet another auxiliary unit vector, $\mathbfit{w}=\mathbfit{a} \boldsymbol\times \mathbfit{N}$, 

\begin{equation}
    w_{x,y,z} = \left\{a_y N_z-a_z N_y,\>  a_z N_x - a_x N_z,\> a_x N_y - a_y N_x\right\}, \label{eq:w}
\end{equation}

\noindent allows us to derive the axis $\mathbfit{b}$ (a unit vector) components in the SSB coordinate system by rotating  the vector 
$\mathbfit{N}$ by the Euler angle $\psi$ towards
the vector $\mathbfit{w}$, with the vector $\mathbfit{a}$ being the rotation vector:

\begin{equation}
    \begin{array}{lcr}
    b_{x,y,z} &=& \left\{N_x\cos\psi + w_x\sin\psi,\>  N_y\cos\psi + w_y\sin\psi,\right.\\
    &&  \left. N_z\cos\psi + w_z\sin\psi\right\}.
    \label{eq:b}
    \end{array}
\end{equation}

\noindent The SSB components for the third axis, $\mathbfit{c}=\mathbfit{a} \boldsymbol\times \mathbfit{b}$, can now be computed as

\begin{equation}
    c_{x,y,z} = \left\{a_y b_z-a_z b_y,\>  a_z b_x - a_x b_z,\> a_x b_y - a_y b_x\right\}. \label{eq:c}
\end{equation}

Brightness models described in \autoref{sec:br_model} require the knowledge of the components of the unit vectors $\mathbfit{S}$ and $\mathbfit{E}$ connecting
the asteroid with the centres of the Sun and Earth, respectively, in the comoving coordinate system ($bca$). These vectors are readily obtainable in the SSB
coordinate system (based on HORIZONS' data). Once the base vectors of the comoving coordinate system, $\mathbfit{b}$, $\mathbfit{c}$, and $\mathbfit{a}$,
have been computed (equations~\ref{eq:a}, \ref{eq:b}, \ref{eq:c}), the components of the vectors $\mathbfit{S}$ and $\mathbfit{E}$ in the $bca$ coordinate system can be calculated as

\begin{equation}
    \begin{array}{lcl}
            S_b &=& b_x S_x + b_y S_y  + b_z S_z,\\
            S_c &=& c_x S_x + c_y S_y  + c_z S_z,\\
            S_a &=& a_x S_x + a_y S_y  + a_z S_z,\\
    \label{eq:S}
    \end{array}
\end{equation}

\noindent and

\begin{equation}
    \begin{array}{lcl}
            E_b &=& b_x E_x + b_y E_y  + b_z E_z,\\
            E_c &=& c_x E_x + c_y E_y  + c_z E_z,\\
            E_a &=& a_x E_x + a_y E_y  + a_z E_z.\\
    \label{eq:E}
    \end{array}
\end{equation}

From the equations listed in this subsection, only the equations~(\ref{eq:XYZ}) are computed once per model integration; 
the rest have to be computed for each observed data point, as they depend on time-variable Euler angles $\theta$, $\varphi$, and $\psi$.

\subsection{Brightness models}
\label{sec:br_model}
\subsubsection{Lommel-Seeliger triaxial ellipsoid}
\label{sec:br_ellipsoid}

We adopt the brightness model of \citet{Muinonen2015a} written for a triaxial ellipsoid with Lommel-Seeliger
light scattering surface. We consider the simplest case of the isotropic single-scattering function.
Disc-integrated absolute magnitude of such an object in our comoving coordinate system $bca$ can be expressed as

\begin{equation}
\setstretch{2}
    \begin{array}{lcl}
    H &=& \Delta V -2.5\log\left\{b' c' \dfrac{T_{\odot} T_{\oplus}}{T} \left[\cos(\lambda'-\alpha') + \cos\lambda' +
            \sin\lambda' \right.\right. \\
       &&    \left.\left. \times \sin(\lambda'-\alpha') \ln\left(\cot\dfrac12 \lambda' \cot\dfrac12(\alpha'-\lambda')\right)\right]\right\},
            \label{eq:V}
    \end{array}
\end{equation}

\noindent were

\begin{equation}
\setstretch{1.5}
    \begin{array}{lcl}
            T_\odot   &=& \sqrt{S_b^2/b'^2 + S_c^2/c'^2 + S_a^2}\quad, \\
            T_\oplus  &=& \sqrt{E_b^2/b'^2 + E_c^2/c'^2 + E_a^2}\quad, \\
            \cos\alpha' &=& (S_b E_b/b'^2 + S_c E_c/c'^2 + S_a E_a) / (T_\odot T_\oplus), \\
            \sin\alpha' &=& \sqrt{1 - \cos\alpha'^2}\quad, \label{eq:Vmisc} \\
            T &=& \sqrt{T_\odot^2 + T_\oplus^2 + 2T_\odot T_\oplus \cos\alpha'}\quad,\\
            \cos\lambda' &=& (T_\odot + T_\oplus\cos\alpha') / T\quad,\\
            \sin\lambda' &=& T_\oplus\sin\alpha' / T
    \end{array}
\end{equation}
           
\noindent  (\citealt{Muinonen2015a}; please note that the authors used $abc$ coordinate system, whereas we use $bca$ coordinate system).
Here $S_{b,c,a}$ and $E_{b,c,a}$ are components of the unit vectors in the directions of the Sun and Earth,
respectively, in the asteroidal coordinate system $bca$ (see Equations~\ref{eq:S} and \ref{eq:E}), $b'$ and $c'$ are the intermediate and smallest semi-axes
of the brightness ellipsoid expressed as a fraction of its largest semi-axis, and the constant $\Delta V$ absorbs two unknown parameters -- albedo and scale (largest semi-axis $a'$ in physical units)
of the asteroid.

Fitting the brightness model (\autoref{eq:V}) to the asteroid's observed light curve, transformed to absolute magnitudes, produces the value
of $\Delta V$ (offset between the model and observed light curves).
To get an estimate of the asteroid's scale $a'_{\rm m}$ (the ellipsoid's largest semi-axis in meters), let us place the asteroid 1~au away from the Earth and Sun, with zero phase angle.
Equation~\ref{eq:V} is then reduced to

\begin{equation}
H=\Delta V-2.5\log (b'c'T_\odot).
\label{eq:H1}
\end{equation}

\noindent As the projected area of the ellipsoid (in square meters) is \citep{Muinonen2015a}

\begin{equation}
A=a'^2_{\rm m} \pi b'c'T_\odot,
\label{eq:A}
\end{equation}

\noindent \autoref{eq:H1} can be rewritten as

\begin{equation}
A=\pi a'^2_{\rm m} 10^{0.4(\Delta V-H)}.
\label{eq:A1}
\end{equation}

\noindent The standard asteroid diameter equation (\citealt{Lamy2004}; their equation~5, written for a spherical body observed at zero phase angle)
can be expressed in terms of the asteroid's projected area $A$ (in square meters) as

\begin{equation}
A=\pi \frac{(1.49598\times 10^{11})^2}{p} 10^{0.4(m_\odot-H)},
\label{eq:A2}
\end{equation}

\noindent where $m_\odot$ is the apparent magnitude of the Sun in the same spectral filter as the one used to observe the asteroid, and
$p$ is the geometric albedo of the asteroid. Equating \autoref{eq:A1} and \autoref{eq:A2} produces the estimate of the physical scale of the ellipsoid:

\begin{equation}
a'_{\rm m} = \frac{1.49598\times 10^{11}}{\sqrt{p}} 10^{0.2(m_\odot-\Delta V)}.
\label{eq:a_m}
\end{equation}

In our code, the triaxial ellipsoid brightness model can be used in two different ways:

\begin{enumerate}
\item Self-consistent case: semi-axes of the brightness ellipsoid, $b'$ and $c'$, are equal to the corresponding 
semi-axes of the kinematic ellipsoid, $b$ and $c$. No additional free parameters.

\item Relaxed case: semi-axes of the brightness ellipsoid are not equal to the corresponding 
semi-axes of the kinematic ellipsoid. Two  additional free parameters: $b'$ and $c'$.
\end{enumerate}

\subsubsection{Black-and-white ball}
\label{sec:BW}

As the simplest case of a non-geometric explanation for the large brightness variations of `Oumuamua, we consider a toy brightness model
consisting of a spherical body with two hemispheres with different albedo values. As `Oumuamua's phase angle is relatively small
($\alpha=19\ldots25^\circ$ in the time interval covered by our analysis), we ignore phase effects for simplicity.

Position of the darker hemisphere (with the albedo described by a free model parameter $\kappa\in[0,1[$) is specified via a 
unit vector $\mathbfit{h}$ (described by two free model parameters: polar angle $\theta_h$ and azimuthal angle $\varphi_h$) in the asteroidal coordinate system
$bca$. The opposite (brighter) hemisphere is considered to have albedo =1.

Ignoring phase effects (phase angle $\alpha=0$), integrated absolute magnitude of such an object is given by

\begin{equation}
    H = \Delta V -2.5\log\left[\kappa\frac{1+\cos\zeta}{2} + \frac{1-\cos\zeta}{2}\right] .
    \label{eq:Vball}
\end{equation}

\noindent Here $\zeta$ is the angle between the vector in the direction of the observer, $\mathbfit{E}$ (see \autoref{eq:E}), and the vector $\mathbfit{h}$.

In total, this brightness model is specified by three free parameters: $\kappa$, $\theta_h$, and $\varphi_h$.

\subsection{Free parameters}
\label{sec:free_params}

Our model can be used with different numbers of free parameters. The most basic model (tumbling self-consistent ellipsoid with zero torque)
has 8 free parameters: $L$ or $P_\psi$, $\theta_L$, $\varphi_L$, $\varphi_0$, $\psi_0$, $E'$, $b$, and $c$. Here $L$ is the modulus of the angular momentum vector,
$P_\psi$ is the rotation period, $\theta_L$ and $\varphi_L$ are the polar and azimuthal angles, respectively, for the angular momentum vector, $\varphi_0$ and $\psi_0$
are the initial values of the precession and rotation Euler angles, respectively,  $E'=2E/L^2$ where $E$ is the rotational kinetic energy, and $b$ and $c$
(used in both kinematic and brightness models) are the intermediate and smallest semi-axes of the ellipsoid expressed in units of the largest semi-axis $a$.

Multiple expanded models (with larger numbers of free parameters) are supported. In particular, relaxing the brightness ellipsoid model
adds two free parameters (brightness ellipsoid's semi-axes $b'$ and $c'$), bringing the total to 10. 
Non-zero torque models add three more free parameters -- normalized torque pseudo vector components in the comoving coordinate system $T_b$, $T_c$, and $T_a$. This
brings the total to 11 and 13 free model parameters, for self-consistent and relaxed brightness ellipsoid models, respectively.

Finally, black-and-white ball brightness model adds three free parameters (dark-to-bright albedo ratio $\kappa$ and polar and azimuthal angles
for the dark side in the comoving coordinate system, $\theta_h$ and $\varphi_h$, respectively) on top of the 8 parameters of the basic model,
bringing the total to 11 free parameters. In the presence of torque, the number grows to 14.

One more parameter, $\Delta V$, is present implicitly (see equations~\ref{eq:V} and \ref{eq:Vball}). It is the offset between the observed and model light curves, in brightness magnitudes.
It is a byproduct of $\chi^2$ fitting of the model brightness curve to the observed one, and can be used to assign physical units to the model (\autoref{eq:a_m}).

Our code can also utilize other free parameters which we did not use for the current project. In particular, our "detrending" parameter $A$ is designed to
approximate the gradual change of the average asteroid's brightness with the phase angle $\alpha$.
Preliminary tests showed that this additional parameter does not improve the quality of fit for our models, which is not surprising given
that we analyse a very short time interval, where `Oumuamua's phase angle changes only slightly (from 19.2 to 24.7$^{\circ}$).

\section{Code}
\label{sec:code}

\subsection{Overview}

We present our code\footnote{The code is publicly available here: \url{https://github.com/pulsar123/Asteroid}}, which fits different models of tumbling asteroids,
described in \autoref{sec:model}, to observed light curves.
It is written in C++ utilizing CUDA framework, and consists of more than 3000 lines of code. With the exception of the brief initialization and finalizing steps, and infrequent 
checkpointing steps, the entire code runs on a GPU as tens of thousands of independent parallel threads, each exploring different optimization paths through 
multi-dimensional free model parameter space. The code runs best on Pascal P100 GPUs (CUDA capability 6.x), but can also be used
on older Tesla GPUs (CUDA capability 2.0 or larger). The massive computational power of modern GPUs makes it realistic to reliably find global minima
in the 8+ dimensional free model parameter space using a Monte-Carlo style optimization strategy. As the optimization engine we use downhill simplex
(Nelder-Mead) method\footnote{\url{https://en.wikipedia.org/wiki/Nelder-Mead_method}} which works well for large number of dimensions and does not require
the knowledge of the partial derivatives of the function to optimize, which in our case is the $\chi^2$ function produced by fitting the model light curve to the observed
one.

Each GPU thread repeatedly goes through the following steps:

\begin{enumerate}
\item A random initial point is generated in the scale-free model parameter space, using CURAND library. The library allows for generation of tens of 
thousands of independent quasi-random number sequences, one for each parallel thread. In the optimization scale-free space, each model parameter is normalized to have 
the initial range of $[0,1]$ (for periodic angle parameters, the $[0,1]$ scale-free range corresponds to the $[0,2\pi]$ radians range). Strongly non-linear
parameters are first linearized. For example, kinematic and brightness ellipsoid semi-axis ratios, $b$, $c$, $b'$, and $c'$, utilize logarithmic scale,
to provide a comparable sampling coverage for each decade of the full parameter's range. Parameter $P_\psi$ (rotation period)
is sampled uniformly in the $1/P_\psi$ (frequency) space.
Some parameters (like $c$, $c'$, and $P_\psi$) have static limits 
for the allowed  range of the initial random values, while other parameters ($b$, $b'$, $\psi_0$, $E'$ etc.) have limits which change during optimization
(they depend on the values of other parameters). The way the initial value of $E'$ is generated is such that SAM and LAM have equal probabilities. 
During optimization, the values of parameters are allowed to drift beyond the initial range ("soft limits"),
as long as they stay within the physically allowed hard limits.

\item The initial simplex is constructed, using a small (typically 0.001 in scale-free units) initial step in each dimension.

\item Every time the optimization ($\chi^2$) function value needs to be computed, the following substeps are performed:

\begin{enumerate}
\item Free model parameters are converted from scale-free to physical units.

\item Initial values of the independent variables in the equations of motions (either \autoref{eq:euler2} or \autoref{eq:euler1}) are computed (\autoref{sec:ICs}).

\item The ODEs (equations of motions; see \autoref{sec:eq_motions}) are solved using the 4th order Runge-Kutta method. 
The integration starts at the time corresponding to the
earliest observed point, and proceeds with steps not larger than $0.01$~d (which in our tests provides sufficient accuracy for `Oumuamua's modeling) from one observed point
to the next one, covering all the observed points. This produces model values of the Euler angles $\theta$, $\varphi$, and $\psi$ 
(plus the values of the angular velocity vector components, $\Omega_b$, $\Omega_c$, and $\Omega_a$, for models with non-zero torque)
for each observed point.

\item For each observed point, the directions of the unit vectors $\mathbfit{S}$ and $\mathbfit{E}$ connecting the asteroid with the centres of the Sun 
and Earth, respectively, are computed via a series of geometric transformations as described in \autoref{sec:coords}.

\item Using one of the brightness models (\autoref{sec:br_model}), the model absolute magnitude of the asteroid is computed for each observed point.
$\chi^2$ value is computed, along with the offset $\Delta V$ between the observed and modeled light curves. Each point
uses the weight of $1/\sigma_i^2$, where $\sigma_i$ is the measurement error for this data point. If the observed data were produced using more than one
spectral filter, separate values of $\Delta V$ are computed for each filter, independently.
\end{enumerate}

\item The downhill simplex algorithm is used to descend to a nearby local $\chi^2$ minimum. The descent is stopped when either the simplex has shrunk below
the smallest allowed size (a sign of being in a local minimum) or a maximum number of simplex steps (typically $5000$) were taken, whichever comes first.
\end{enumerate}

\subsection{Numerical runs}
\label{sec:runs}

We tried several multi-stage optimization strategies. The one we adopted performed the best overall in our validation tests (see below), and consists of the
following steps:

\begin{enumerate}
\item {\it Random search stage.} Eight instances of the code is run on eight P100 GPUs for 24 hours. There are $\sim 30,000$ 
threads running in parallel on each GPU. Each GPU thread starts at a random
point in the free model parameter space. Once all the threads in a block of 256 threads converge to local $\chi^2$ minima or hit the 
simplex step limit, the best model (lowest $\chi^2$) of the block is chosen. It is then used to seed the second search phase
(searching in the neighborhood of the best model), where the same 256 threads would start at points which are randomly and slightly offset from the best model,
with randomized initial simplex steps for each dimension. The second phase ends using the same criteria: all the block threads either converge to local minima
or hit the simplex step limit. The best model of the block in the second phase is stored in a file.  At the end, about $10^5$ best models are written to files.

\item {\it Re-optimization stage.} Eight instances of the code running on eight P100 GPUs for 24 hours is launched after the first 
stage is completed. The GPU farm goes through the sorted list of the models found in the first stage, starting from the best (smallest $\chi^2$) model.
Each code instance searches in the neighborhood of a model from the first stage in multiple (typically 20) global re-optimization steps. Each global step
consists of all parallel threads on a GPU (typically $\sim30,000$) searching for the best $\chi^2$ minimum in the neighborhood of the globally best model found
by all the parallel threads in the previous step. Around 1~per cent of the best models from Stage One are processed in Stage Two, resulting in $\sim 10^3$ highly
optimized models.

\item {\it Fine-tuning stage.} Finally, a few best models from Stage Two are subjected to additional global re-optimization steps, using higher (double, vs. single in prior
stages) floating point precision and much smaller minimum simplex diameter ($10^{-10}$ vs. $10^{-5}$, in scale-free units), until a numerical convergence is achieved.
For models where some light curve minima still have an obvious offset from the observed ones in the time dimension, an attempt to drive all the major model minima 
towards nearest observed ones is made. This is achieved by means of progressively decreasing the optimization ($\chi^2$) function value as the model minima
converge to the observed ones in the time dimension.
\end{enumerate}

\subsection{Code validation: artificial dataset test}
\label{sec:fake}

\begin{figure}
   \includegraphics*[width=\columnwidth]{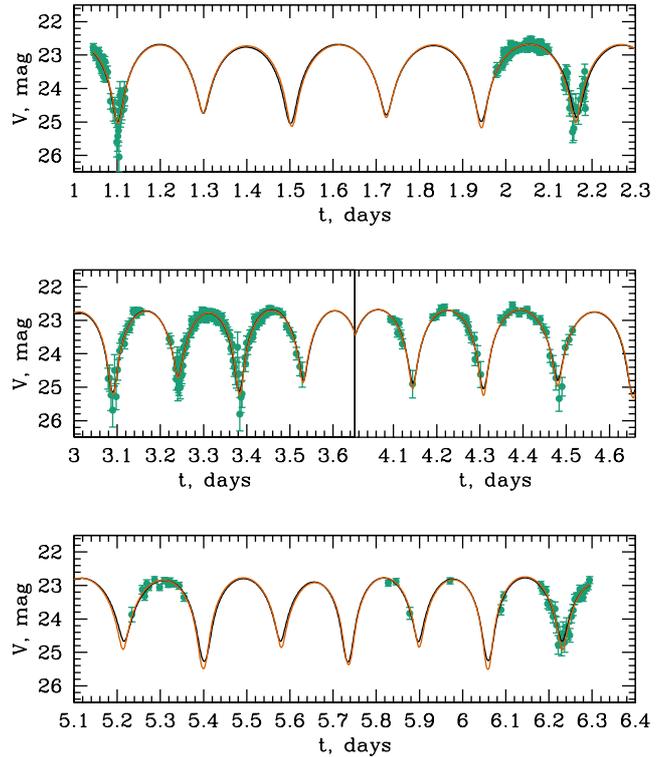}
    \caption{Artificial dataset test. $t$ is the number of days since
    ${\rm MJD}=58050$, and $V$ is the absolute magnitude in the $r$ spectral filter. 
    Artificial data points with one-sigma errorbars are shown in green (grey in the printed version of the journal). Black line
    depicts the light curve of the underlying model. Orange (grey in the printed version of the journal) line corresponds to the best-fitting model. 
     }
    \label{fig:fake}
\end{figure}

\begin{table*}
\caption{Models.}
\label{tab:models}
\begin{tabular}{lcccccccccc}
\hline
Parameter                & Fake\_ini    & Fake\_fit  &TD60$_{\rm A}$&TD60$_{\rm B}$&TD60$_{\rm C}$&  INERT  &  DISC      &  CIGAR     & SAIL       & BALL     \\
\hline
\multicolumn{11}{c}{\rule{0pt}{3ex}\it Fitting parameters}\\
$\chi^2$                 &     -        & 1.011      & 6.275        & 2.283        & 4.135        & 18.40   & 9.963      & 10.84      & 11.50      & 7.859    \\       
rms, mag                 &     -        & 0.116      & 0.124        & 0.075        & 0.101        & 0.299   & 0.220      & 0.230      & 0.236      & 0.195    \\
$\Delta V$, mag          &   22.6720    & 22.5066    & 18.8774      & 18.3002      & 18.2139      & 19.3378 & 22.5287    & 20.2872    & 22.0703    & 22.5523  \\
\multicolumn{11}{c}{\rule{0pt}{3ex}\it Free model parameters}\\
$\theta_L$, rad          &   2.23573    & 1.05433    & 0.97459      & 1.91359      & 2.50770      & 0.90522 & 2.27661    & 2.02359    & 1.42107    & 1.57272  \\
$\varphi_L$, rad         &   1.62370    & 4.96208    & 5.82597      & 2.68432      & 2.68163      & 3.00312 & 1.61185    & 1.96810    & 2.36917    & 4.67914  \\
$\varphi_0$, rad         &   6.25839    & 3.94604    & 2.91895      & 0.08683      & 0.17613      & 5.12912 & 6.24557    & 2.96016    & 3.23746    & 4.58948  \\
$T_b$                    &   $-$4.02658 & $-$4.08395 &    -         &    -         & -            & -       & $-$3.97579 & 2.55650    & 47.8356    & 15.4057  \\
$T_c$                    &   $-$1.11909 & $-$1.04747 &    -         &    -         & -            & -       & $-$1.12470 & $-$7.43845 & 0.40040    & $-$2.23678 \\
$T_a$                    &   $-$5.85003 & $-$5.22063 &    -         &    -         & -            & -       & $-$5.82852 & 1.01262    & $-$21.4239 & 5.75293  \\
$c$                      &   0.16435    & 0.18766    & 0.56295      & 0.52170      & 0.53268      & 0.00999 & 0.16293    & 0.12972    & 0.00001    & 0.52318  \\
$b$                      &   0.96483    & 0.96332    & 0.64739      & 0.78211      & 0.80860      & 0.06014 & 0.96427    & 0.13144    & 0.80683    & 0.93090  \\
$E'_0$                   &   0.98270    & 0.97923    & 0.65202      & 0.58023      & 0.60140      & 0.00820 & 0.98204    & 0.03353    & 0.42320    & 0.86537  \\
$L_0$                    &   17.0921    & 16.9094    & 97.9840      & 93.5938      & 90.4275      & 293.884 & 17.0497    & 525.649    & 51.4673    & 42.4590  \\
$\psi_0$, rad            &   1.97276    & 1.91811    & 1.70236      & $-$0.48206   & $-$0.28808   & 6.14166 & 1.99745    & 0.08955    & 0.07324    & 0.52839  \\
$c'$                     &   0.17488    & 0.14139    & 0.31844      & 0.20757      & 0.19353      & 0.01375 & -          & -          & -          & -        \\
$b'$                     &   0.99941    & 0.87302    & 0.67547      & 0.78147      & 0.76291      & 0.06346 & -          & -          & -          & -        \\
$\theta_h$               &   -          & -          & -            & -            & -            & -       & -          & -          & -          & 2.93279  \\
$\varphi_h$              &   -          & -          & -            & -            & -            & -       & -          & -          & -          & 1.96926  \\
$\kappa$                 &   -          & -          & -            & -            & -            & -       & -          & -          & -          & 0.03083  \\
\multicolumn{11}{c}{\rule{0pt}{3ex}\it Derived parameters}\\
$P_{\psi,0}$, h          &   52.01      & 52.70      & 6.784        & 6.783        & 6.788        & 7.670   & 51.81      & 80.84      & 7.750      & 23.92 \\
$P_{\varphi,0}$, h       &   10.75      & 10.89      & 2.850        & 2.852        & 2.851        & 138.3   & 10.79      & 8.557      & 7.163      & 4.509 \\
MODE$_0$                 &   LAM        & LAM        & LAM          & SAM          & SAM          & LAM     & LAM        & SAM        & SAM        & SAM \\
\\
$P_{\psi,1}$, h          &   32.37      & 33.98      &    -         &    -         &   -          & -       & 32.28      & 29.45      & 8.194      & 20.12 \\
$P_{\varphi,1}$, h       &   10.83      & 10.52      &    -         &    -         &   -          & -       & 10.81      & 8.895      & 7.685      & 3.930 \\
MODE$_1$                 &   SAM        & SAM        &    -         &    -         &   -          & -       & SAM        & LAM        & SAM        & LAM \\
\hline                                                            
\end{tabular}                                                     

\justify
{\it Note.} Different columns correspond to different models. The units for free model parameters are set by $a=1$ and $I_a=1$; the time unit is a day. For models with torque, 
$E'_0$ and $L_0$ values are the initial values, for zero torque models they are fixed in time. For the derived parameters $P_\psi$, $P_\varphi$, 
and MODE,
both the initial (subscript 0) and final (subscript 1; only for models with torque) values are provided.
Models TD60$_{\rm A}$ and TD60$_{\rm B}$ are for the second half of the full dataset for 2002 TD$_{60}$ (993 points);
model TD60$_{\rm C}$ is for the full dataset (1914 points). For the three 2002 TD$_{60}$ models, only one value of $\Delta V$ (corresponding to the 
observations calibrated to $R$ filter) is provided. In $\chi^2$ computations, we arbitrarily assumed the 2002 TD$_{60}$ brightness measurements to have 
the std of $0.05$~mag. The initial MJD moments of time in the asteroidal coordinate system are 58053.31317 (all Fake and `Oumuamua models),
52609.73185 (TD60$_{\rm A}$ and TD60$_{\rm B}$ models), and 52585.01840 (TD60$_{\rm C}$ model).
\end{table*}

We performed two different kinds of validation tests with our code. The first one, described in this subsection, is designed to test the
internal self-consistency of the code, using fake light curve data which were created to be as close to the observed light curve data for `Oumuamua as possible.

To generate the fake data, we used one of our best-fitting models for `Oumuamua (see \autoref{sec:data}) which consists of a 
relaxed brightness ellipsoid
model and a tumbling rotation model subject to fixed torque (13 free model parameters in total; see \autoref{sec:free_params}). This model is for a
thin (1:6 ratio) disc-like object which is close to being self-consistent (brightness ellipsoid semi-axes $b'$ and $c'$ are close to the corresponding
kinematic ellipsoid semi-axes, $b$ and $c$). To generate fake light curve data based on
this model, we computed the model absolute magnitudes for the same number of observations at exactly the same moments of time as in `Oumuamua's dataset.
Next we degraded the data using Gaussian noise with the std following the same trend as the observational uncertainties:
$\sigma \approx 10^{0.323 H-8.416}$. (Here $H$ is the
absolute magnitude of the asteroid, and units for $\sigma$ are magnitudes.)
The same values of $\sigma$ were later used as fake observational uncertainties during $\chi^2$ model fitting. 
\autoref{fig:fake} shows the fake data as dots with errorbars, and the underlying model as a black curve.

We next performed a full set of model numerical runs (as described in \autoref{sec:runs}) on the fake data, using the same optimizations parameters
and the same soft limits for free model parameters as in `Oumuamua runs (\autoref{sec:torque}). The free model parameters were the same 13 parameters
used to generate the underlying model. At the end we did find a few very good model matches; the best one is shown as an orange (grey in the printed version of the journal) curve in \autoref{fig:fake}.
Despite the significant noise present in the data and its patchiness, the recovered
model is very close to the underlying model.

Full details of the underlying and best-fitting models are listed in \autoref{tab:models} in Fake\_ini and Fake\_fit columns, respectively. The only obvious
difference between the two models is the opposite direction of the angular momentum vector, which affected the angles $\theta_L$,
$\varphi_L$, and $\varphi_0$. The rest of the recovered parameters are close to their original values, including the three normalized torque vector components,
$T_b$, $T_c$, and $T_a$. The $\chi^2$ value for the best-fitting model ($1.011$) is close to one, as expected.

The artificial dataset test demonstrated the following:

\begin{enumerate}
\item The code is internally self-consistent (it can recover its own models from noisy data).
\item The underlying model can be recovered for models as complex as the one used for testing
(13 free model parameters, including three torque parameters), and for data as patchy and noisy as `Oumuamua's light curve dataset.
\end{enumerate}

\subsection{Code validation: tumbling asteroid 2002 TD$_{60}$}

The artificial dataset test described in the previous subsection validated many important aspects of our code and numerical procedure, but it lacks
physical validation of the code and model. (Being internally self-consistent does not mean the model is correct and physical.) We addressed this shortcoming
by using our code to recover the parameters of a well studied tumbling asteroid, 2002 TD$_{60}$ \citep{Pravec2005}.

The Amor asteroid 2002 TD$_{60}$ is one of the best studied NPA rotators. It was observed on multiple telescopes during the observational campaign
in November--December 2002, producing
a significant number (1914) of high accuracy measurements of the asteroid's brightness \citep{Pravec2005}. Around 30~per cent of the measurements (544 points) 
are calibrated ($R$ spectral filter); the rest consist of eight uncalibrated subsets. This necessitates the use of nine independent fitting parameters $\Delta V$ 
(one for each internally self-consistent subset of the data) when fitting a  model to the full dataset.

\citet{Pravec2005} obtained good model fits to the light curve of 2002 TD$_{60}$. In many respects their approach is similar to ours; in particular, their 
brightness model is a triaxial ellipsoid with LS reflectance law, and the optimization engine is the simplex downhill method. But there are some non-trivial
differences. Importantly, their brightness model is a numerical one (the triaxial ellipsoid is represented by 2292 flat triangles), whereas we use a more accurate
and reliable (and much faster to compute) analytical formulation of \citet{Muinonen2015a}. (One has to note that the numerical approach is more
flexible as one can easily modify the reflectance law.) Another significant difference is the fact that \citet{Pravec2005} had to estimate many of 
the model parameters (main frequencies, brightness ellipsoid axes ratio) before performing the simplex downhill optimization, whereas we employ a brute 
force optimization approach in which no model parameter estimates are used. The brute force approach is more advantageous as it explores the whole free model parameter space,
including the regions which may be overlooked in a constrained approach. Our approach was made possible by the dramatically faster computing hardware available today,
and also thanks to the brightness model being analytical.

\begin{figure}
    \includegraphics*[width=\columnwidth]{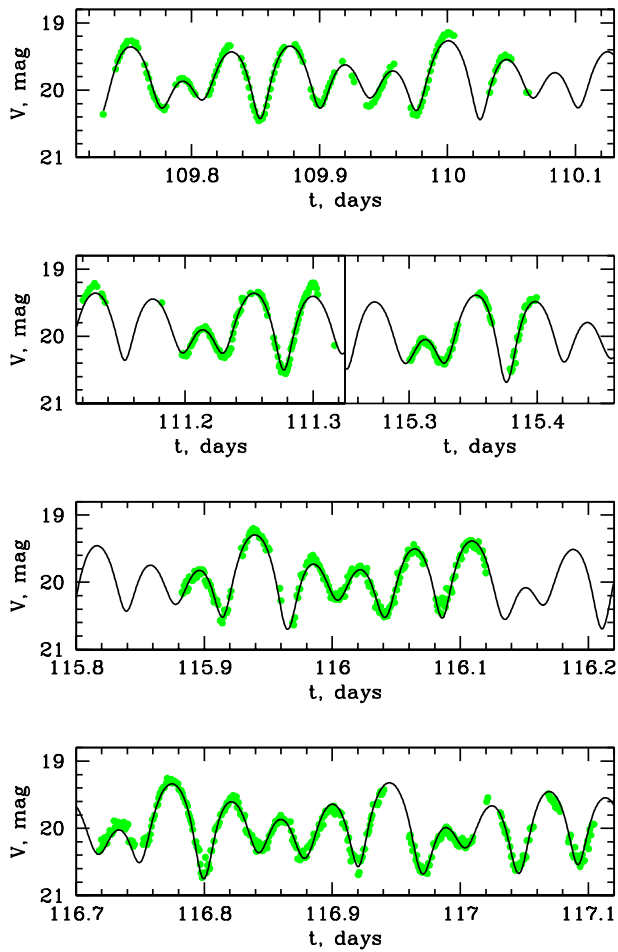}
    \caption{Our best-fitting tumbling asteroid model (TD60$_{\rm B}$, see \autoref{tab:models}) for the asteroid 2002 TD$_{60}$. $t$ is the number of days since
    ${\rm MJD}=52500$, and $V$ is the absolute magnitude in the $R$ spectral filter.
    Dots are the 993 observational data points from \citet{Pravec2005}.
    The top and bottom panels
    correspond to the panels (c) and (d) of figure~4 of \citet{Pravec2005}.}
    \label{fig:TD60}
\end{figure}

We carried out a full suite of numerical runs, as described in \autoref{sec:runs}, to fit our model to the light curve data for 2002 TD$_{60}$ 
(generously provided by the author; P. Pravec 2018, private communication). Most of our analysis was restricted to the second half of the full dataset 
(993 out of 1914 points; ${\rm MJD}=52609.7$--$52617.1$; five independent $\Delta V$ parameters). As TD$_{60}$ observations span a significantly longer
time interval (11--35~d) than `Oumuamua's dataset (5~d), we had to use a shorter ODE integration time step (0.005~d vs. 0.01~d) to achieve
a good numerical convergence.
As \autoref{fig:TD60} shows, our best-fitting model fits very well the observed light curve for the asteroid. Full details
for this model are listed in the TD60$_{\rm B}$ column of \autoref{tab:models}.
The quality of the fit is substantially better than for the best-fitting model of \citet{Pravec2005}: the rms is 0.075 and 0.18~mag, respectively.
We should caution that these rms values are for different subsets of the dataset. To make a more meaningful comparison, we re-optimized our model TD60$_{\rm B}$ 
for the full dataset (1914 points), which produced a slightly worse fit,
with the rms of 0.10~mag (model TD60$_{\rm C}$ in \autoref{tab:models}) -- still almost factor of two better than the best-fitting model of \citet{Pravec2005}.

Interestingly, our best-fitting model is substantially different from the one derived by \citet{Pravec2005}. The rotation and precession periods are essentially
identical (our model: $P_\psi=6.783$~h and $P_\varphi=2.852$~h; model of \citealt{Pravec2005}: $P_\psi=6.787$~h and $P_\varphi=2.851$~h),
but the models are rather different otherwise. Importantly, our best-fitting model is in a SAM tumbling motion, whereas the model of \citet{Pravec2005} is
a LAM rotator.

To get more clarity, we used our code to search for a best-fitting model in the neighbourhood of the best-fitting model of \citet{Pravec2005}, using the same dataset
as for our model TD60$_{\rm B}$ (993 points). We did find a local $\chi^2$ minimum which corresponds to a model (column TD60$_{\rm A}$ in \autoref{tab:models}) which is much
closer to the best-fitting model of \citet{Pravec2005}. Importantly, both models are LAM rotators. The periods are again almost identical, but now other
model parameters are close as well: $c=\{0.56,0.54\}$, $b=\{0.65,0.70\}$, $c'=\{0.32,0.36\}$, $b'=\{0.68,0.64\}$ (for our model and that of \citealt{Pravec2005},
respectively; using our notation). The rms for TD60$_{\rm A}$ is $0.124$~mag, which is $1.7$ times larger than for our globally best-fitting model, TD60$_{\rm B}$.

Our light curve model fitting for the tumbling asteroid 2002 TD$_{60}$ appears to be largely consistent with the previously published
results \citep{Pravec2005}. The small differences in the best-fitting model parameters can likely be attributed to numerical \citep{Pravec2005} versus analytical
(present paper) implementations of the LS triaxial ellipsoid integrated brightness model. The biggest factor appears to be our brute force optimization
approach which allowed us to thoroughly explore the free model parameter space and as a consequence to discover a substantially better solution.

At the end, the second (physical) code validation test proved to be not as clear-cut as we hoped. Nevertheless, we did gain extra confidence in the code and model
correctness, and again confirmed the code's power to discover good model fits for realistic (noisy and patchy) light curve datasets, for asteroids
which shape is not exactly a triaxial ellipsoid.

\section{Modeling `Oumuamua's light curve}
\label{sec:oumuamua}

\subsection{Observational data}
\label{sec:data}

\begin{figure}
    \includegraphics*[width=\columnwidth]{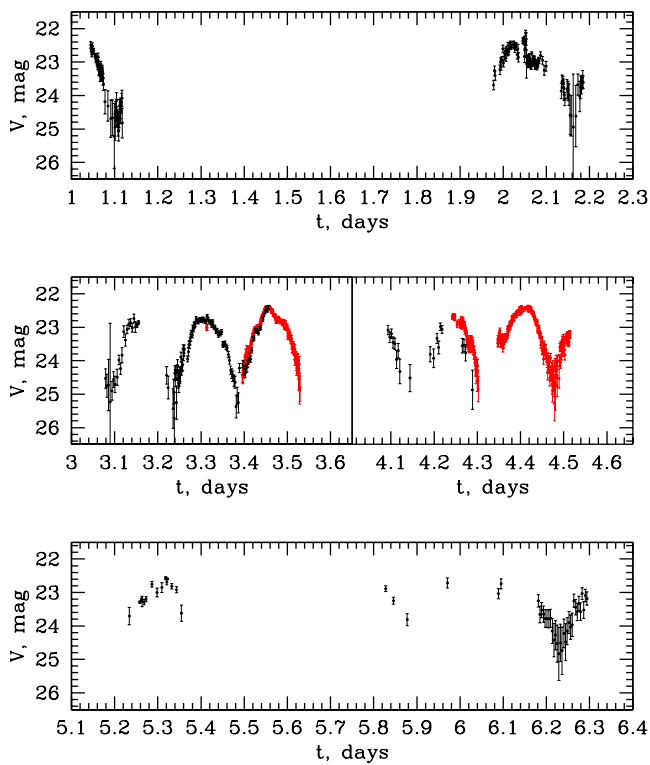}
    \caption{Observed light curve for `Oumuamua (770 points) based on the datasets of \citet[black points]{Fraser2018} and \citet[red points; grey points in the printed version of the journal]{Drahus2018}.
    $t$ is the number of days since
    ${\rm MJD}=58050$, and $V$ is the absolute magnitude in the $r$/$r'$ spectral filter.  The errorbars
    are one-sigma uncertainties.  }
    \label{fig:data}
\end{figure}

We used two sources of `Oumuamua's light curve data, together covering five days of observations, from ${\rm MJD}=58051.0$ (October 25, 2017) to ${\rm MJD}=58056.3$ 
(October 30, 2017). This is the only time interval when the asteroid's brightness 
measurements were very frequent (but still rather patchy, with large gaps between individual observational runs; see \autoref{fig:data}),
resulting in reliable detection of multiple features (minima and maxima). The few other existing observations excluded from our analysis \citep{Belton2018}
are very sparse, lack any obvious features
(significantly reducing their value for model fitting), and span a much longer time interval (one month), which would make model computations a factor of five
slower. Over the course of these five days, the asteroid--Sun distance ranged from 1.36 to 1.49~au, the asteroid--Earth distance ranged from
0.40 to 0.58~au, and the phase (Earth -- asteroid -- Sun) angle ranged from 19.2 to 24.7$^{\circ}$. The small range of the phase angle change is particularly
helpful, as it minimizes the need for sophisticated light scattering formulations in our brightness models.

The first source of data \citep{Fraser2018} is itself a compilation of optical photometry of `Oumuamua from multiple publications 
\citep{Meech2017,Bannister2017,Jewitt2017,Bolin2018,Knight2017}. It consists of 339 `Oumuamua brightness measurements, converted to the same spectral filter
($r'$) using known spectral properties of the asteroid. Some of these observations have very large one-sigma uncertainties; 
the full range is $0.02 \ldots 2.4$~mag, with the geometric mean of 0.19~mag. The dataset of \citet{Fraser2018} is corrected for light travel 
(times correspond to the asteroidal coordinate system), and converted to absolute magnitudes (corresponding to both asteroid--Earth and asteroid--Sun distances
of 1~au).

The second source of data is a homogeneous set of 431 high quality `Oumuamua's brightness measurements carried out  in the $r$ spectral band
using Gemini Multi-Object Spectrograph (GMOS-N) over the course of two nights, October 27--28, 2018
\citep{Drahus2017,Drahus2018}. The times are light travel corrected to the standard epoch of
${\rm MJD}=58054$; we had to subtract $0.0028662$~d from all times to convert them to the asteroidal coordinate system. 
The brightness values are geometry corrected to 
${\rm MJD}=58054$; to convert them to absolute magnitudes we had to add $0.74$~mag to the published values. One-sigma brightness measurements
uncertainties range from 0.02 to 0.5~mag, with the geometric mean of 0.08~mag. The full dataset is not publicly available, but was generously provided
by the authors (Michal Drahus 2018, private communication). The full dataset only became available recently, in the final stages of this research project.
Most of our numerical runs used a shorter version of the dataset, consisting of 51 points manually scanned from figure~4 of \citet{Drahus2017}. Once
the full dataset became available, we made sure that our reduced dataset is fully consistent with the actual data. The final re-optimization steps
in our analysis presented here are based on the full dataset (431 points).

Merging the two datasets together produced a list of either 390 (when using the scanned version of \citealt{Drahus2017} dataset) or 770 (when using the
full dataset of \citealt{Drahus2018}) `Oumuamua's absolute magnitude values in the $r$/$r'$ spectral filter with the corresponding one-sigma uncertainties, with the
times corrected for light travel. These are the data we used in all our light curve modeling efforts for this asteroid. The data are plotted in
\autoref{fig:data} as black and red (grey in the printed version of the journal) points with one-sigma errorbars.

As noted by many authors before, the most striking feature of `Oumuamua's light curve is the presence of multiple very deep minima (see \autoref{fig:data}). 
Some of the minima are defined by few
points with large measurement errors so may not appear very significant individually (for example, minima A, B, I, and L in \autoref{fig:inert}),
but taken together they present a very convincing case for an object undergoing
extreme brightness variations (with the amplitude up to 2.5--2.6~mag -- larger than any known Solar System asteroid, \citealt{Jewitt2017}) on a quasi-regular basis.
The conventional interpretation of these brightness variations is that they are caused by extremely elongated (if it is cigar-like) or flattened 
(if it is disc-like) shape of `Oumuamua \citep{Meech2017}, though non-geometric interpretations (e.g. extreme albedo variations across the object's
surface) cannot be ruled out. Assuming the geometric interpretation and ignoring phase effects, the light curve amplitude of 2.5~mag would correspond
to the cigar or disc largest-to-smallest axes ratio of 10:1 \citep{Meech2017}. When taking into account phase effects, the shape constraints are
not as extreme (>5:1, \citealt{Fraser2018}), though still quite remarkable.

\subsection{Inertial ellipsoid models}
\label{sec:inertial}

\begin{figure}
    \includegraphics*[width=\columnwidth]{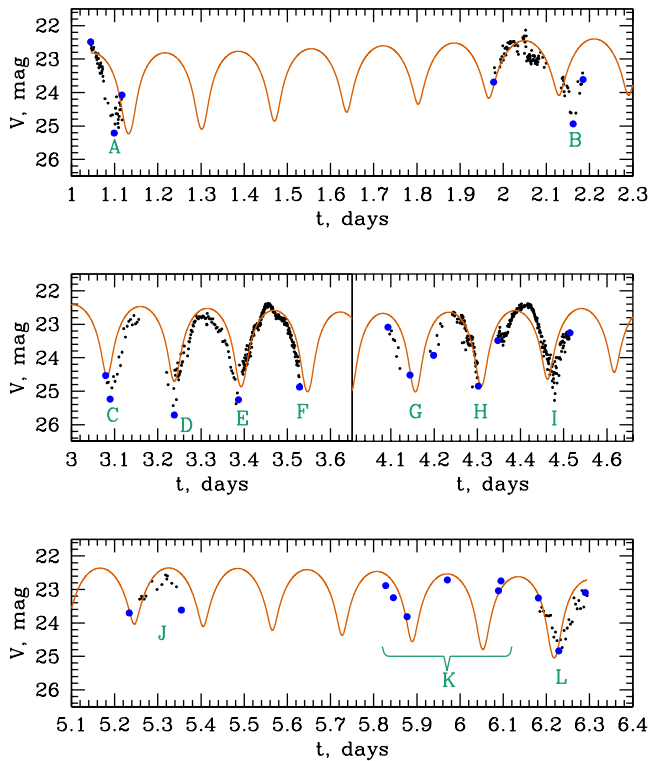}
    \caption{Our best-fitting inertial model (column INERT in \autoref{tab:models}) for `Oumuamua.
    $t$ is the number of days since
    ${\rm MJD}=58050$, and $V$ is the absolute magnitude in the $r$/$r'$ spectral filter.
    The dots are the observational data (errorbars were omitted to make trends more obvious). Blue dots (open circles in the printed version of the journal) are the anchor points
    with the std reduced to $0.05$~mag, which purpose is to define major features (primarily deep minima) in the observational data. 
    Capital letters A--L mark the positions of major features in the observational data.
    }
    \label{fig:inert}
\end{figure}

The first class of models we used to simulate the light curve of `Oumuamua is a 10-parameter inertial (zero torque) 
relaxed LS triaxial ellipsoid model (see \autoref{sec:free_params}). The 10 free model parameters
had the following soft (hard) limits: $\theta_L \in [0,\pi[$ (same), $\varphi_L \in [0,2\pi]$ (any),
$\varphi_0 \in [0,2\pi]$ (any), $c \in [0.01,1[$ ($]0,1[$), $b \in [c,1[$ (same), $E' \in [0,1]$ (same), $P_\psi \in [2,4800]$~h
($P_\psi > 0$), $\psi_0 \in [\psi_{\rm min},\psi_{\rm max}]$ for SAM (same; see \autoref{eq:psi}) and $\psi_0 \in [0,2\pi]$ for LAM (any),
$c' = c$ ($]0,1[$), $b' = b$ ($[c',1[$). (Soft limits are used when generating initial random values of the parameter; hard limits
are enforced during optimization; see \autoref{sec:free_params} for the explanation of the parameters.) 

Early attempts of model fitting produced completely unsatisfactory results, with hundreds of lowest $\chi^2$ models failing to reproduce
the quantity and locations of the major features (minima and maxima) of the observed light curve. This did not occur in our artificial 
dataset tests (\autoref{sec:fake}), despite the fact that the fake data were as noisy and patchy as `Oumuamua's data, and
the "fake" model being more complex (three additional model parameters -- the three normalized torque vector components). This was an early
indication that the inertial ellipsoid model was not the right one for this asteroid. To try to match at least the quantity and locations
of the obvious observed minima, we started to reduce the std for some of the data points, which define the major light curve features,
to a small value of 0.05~mag (comparable to the best real std in the data). At the end of this rather lengthy iterative process
we ended up with 28 "anchor" points (see \autoref{fig:inert}). (It is important to note that we only used the "anchor" points
with fake std values during the stages one and two of our optimization procedure; during the final -- fine-tuning -- stage we
used correct std values for all the data points.)

We also employed another trick during the fine-tuning stage (\autoref{sec:runs}), when the model minima located close to the seven
most obvious observed minima (features A, B, C, D, E, I, and L in \autoref{fig:inert}) would be gradually "nudged" towards the
corresponding observed minima in the time dimension. This is accomplished by multiplying the $\chi^2$ values by a parameter
$\beta$ which is equal to one when the minima are far apart, and becomes significantly smaller than one when all seven
model minima converge onto the corresponding observed minima.

The above tricks allowed us to produce somewhat better model fits. But even the best of them (our model INERT; see \autoref{fig:inert}
and \autoref{tab:models}) was completely unsatisfactory: some of the model minima (especially the feature B, also A and L) 
were offset in the time dimension from the corresponding observed minima by a non-trivial amount. 
The fact that relaxing of the brightness ellipsoid model (by means of adding two more free model parameters) failed to produce
models which would be at least in a qualitative agreement with the observed light curve is highly suggestive of significant issues
with the current model. The possible explanations for the model failures are:

\begin{enumerate}
\item The shape of `Oumuamua is dramatically different from the assumed triaxial ellipsoid shape. This explanation cannot be ruled out based on our analysis,
but we consider it to be very unlikely. As our 2002 TD$_{60}$ test case shows, the triaxial ellipsoid brightness model has no issues in fitting 
the observed minima of a real (that is, not with a perfect triaxial ellipsoid shape) asteroid in the time dimension.
It does not do as well in terms of explaining the detailed shape of minima and maxima, but relaxing
the brightness model (by means of adding two more free parameters, $c'$ and $b'$)
improves the quality of the fit substantially, to a large degree taking care of the shape mismatch. 
(This is fully consistent with much more extensive results of \citealt{Cellino2009} and \citealt{Muinonen2015b}.)
A significantly different
reflection law would also unlikely fix the significant offsets between the model and observed light curve minima in the time dimension \citep{Samarasinha2015}.

\item In light of the discovery of `Oumuamua's non-gravitational acceleration \citep{Micheli2018}, a natural expansion of our model
would be to assume a presence of some torque. We consider the simplest prescription for torque (fixed in time and spatially, in the asteroidal coordinate system)
in our ellipsoid models with torque simulations (see \autoref{sec:torque}).

\item The large brightness variations of `Oumuamua are not geometric in nature (caused by extreme shape of the object). An alternative explanation
would be an object with a more conventional shape (say, roughly spherical), but with extreme albedo variations across the surface. We explore
this alternative explanation via our "black-and-white ball" model (see \autoref{sec:alternative}).

\end{enumerate}

\subsection{Ellipsoid models with torque}
\label{sec:torque}

\begin{figure}
    \includegraphics*[width=\columnwidth]{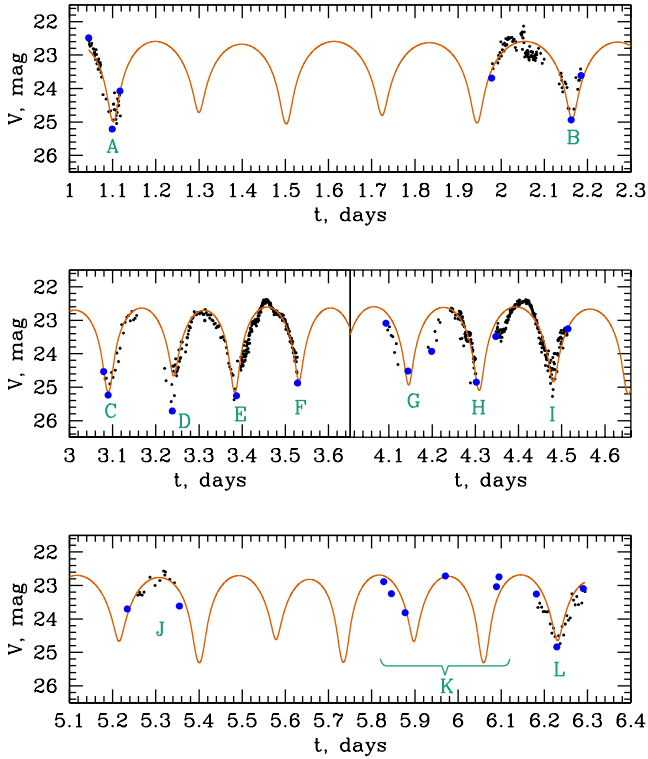}
    \caption{Our best-fitting disc model with torque (column DISC in \autoref{tab:models}) for `Oumuamua.
    See the caption of \autoref{fig:inert} for details.
    }
    \label{fig:disc}
\end{figure}

\begin{figure}
    \includegraphics*[width=\columnwidth]{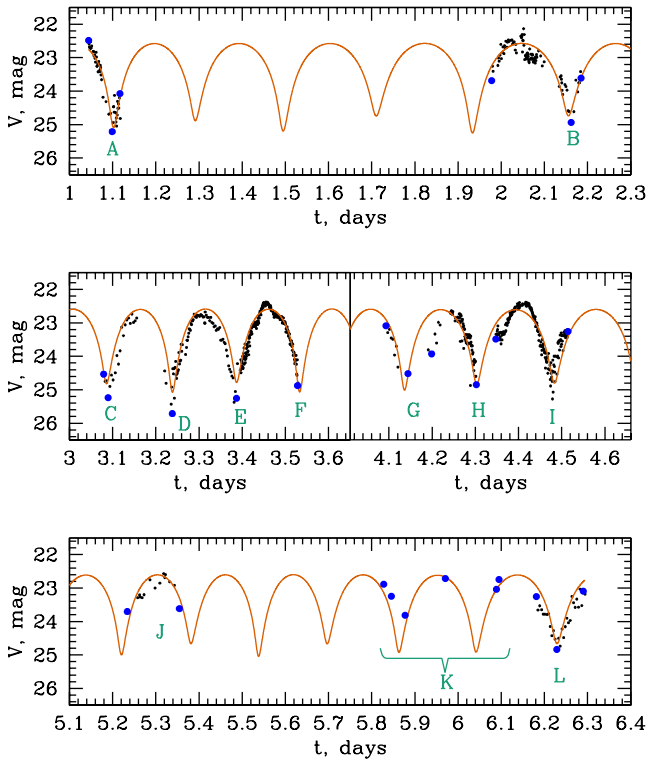}
    \caption{Our best-fitting cigar model with torque (column CIGAR in \autoref{tab:models}) for `Oumuamua.
    See the caption of \autoref{fig:inert} for details.
    }
    \label{fig:cigar}
\end{figure}

The most obvious (in light of the detected non-gravitational acceleration of `Oumuamua, \citealt{Micheli2018}) 
and simplest extension to our inertial tumbling rotation model is to add constant torque (fixed in the asteroidal coordinate system).
This adds three more free parameters (normalized torque vector components $T_b$, $T_c$, and $T_a$), and doubles the number of ODEs (from 3 to 6) in the equations of motion
(\autoref{eq:euler2}). 

We carried out the standard suite of numerical runs (\autoref{sec:runs}) to find best-fitting `Oumuamua models for both self-consistent
and relaxed LS ellipsoid brightness models (11 and 13 free model parameters, respectively). We used the same soft and hard limits for the basic
model parameters as in our inertial model runs (\autoref{sec:inertial}); for the three components of the normalized torque vector we ended up using
the soft limits $[-10,10]$ (no hard limits) in the model physical units (where $a=1$ and $I_a=1$; the time unit is a day). Preliminary tests showed
that with a factor of 10 larger soft limits the vast majority of best-fitting models end up spinning unphysically fast (periods less than one hour)
at the end of the 5-day simulated time interval, producing light curves which looked totally wrong. A factor of 10 smaller soft limits, $[-1,1]$,
produced best-fitting models similar to our best-fitting inertial models, suggesting that in these models torque was too weak to make an obvious impact.

In what we consider to be the main result of this paper, we found that adding the simplest (constant) torque prescription to the inertial tumbling asteroid model
significantly improves the quality of model fits to `Oumuamua's light curve. Importantly, the timings of the well defined
and sharp observed brightness minima can now be matched very well by the models (Figures~\ref{fig:disc} and \ref{fig:cigar}).
In both relaxed and self-consistent brightness ellipsoid runs we identified
two classes of models which were consistently in the top 5--10 best models in terms of the lowest $\chi^2$ values, had model minima matching well
the timings of the observed minima, and also reproduced well other major features of the observed light curve (for example features J and K).

The first class of models has the best overall $\chi^2$ values, and is comprised of thin discs which are almost self-consistent and close to being axially symmetric.
An interesting point is that making the brightness model relaxed (by adding two shape parameters -- $b'$ and $c'$) does not improve the quality of fit
(in terms of $\chi^2$, rms, and matching the timings of the observed minima) for this class of models. As a result, here we present only the self-consistent
version of this model (\autoref{fig:disc}; column DISC in \autoref{tab:models}). As one can see, it is still far from being perfect. In particular,
the large observed depth of the minimum D is not correctly reproduced, the shape of the minimum G is not well matched, and the model feature L is systematically
raised relative to the observed one (\autoref{fig:disc}). Also, the $\chi^2$ and rms values (10.0 and 0.22~mag, respectively) are still fairly large,
albeit much smaller than for our best-fitting inertial model INERT (18.4 and 0.30~mag, respectively; see \autoref{tab:models}). Despite all of this,
the ability of our torque models to match all the main features of the observed light curve of `Oumuamua is quite remarkable. The remaining deviations
of the model light curve from the observed one can be plausibly ascribed to non-ellipsoidal shape, non-homogeneous albedo, and/or more complex light scattering 
properties of the asteroid.

Our best-fitting model DISC (\autoref{tab:models}) is a thin ($1:6.1$) disc which is very close to being axially symmetric ($b/a=0.96$) and 
which is in a LAM rotation initially (with the rotation and precession periods 51.8 and 10.8~h, respectively). It is interesting that by the end of
the simulated time interval of 5 days, constant torque turns the asteroid into a SAM rotator, reducing the rotation period by a factor of 1.6, but
keeping the precession period essentially unchanged. Assuming geometric albedo $p=0.1$ and adopting the Sun's visual magnitude in $r$ filter
$m_\odot=-27.04$~mag from \citet{Willmer2018}, we estimate the physical dimensions (full diameters) of the model
as $115\times 111\times 19$~meters  (from \autoref{eq:a_m}).

The second class of best-fitting models has slightly larger (which is likely statistically insignificant) values of $\chi^2$ and rms,
and is comprised of narrow ($1:7.7$) cigar-shaped objects which are close to being axially symmetric ($c/b=0.99$).
The best representative of this class -- model CIGAR (\autoref{fig:cigar}, \autoref{tab:models}) -- has comparable $\chi^2$ and rms values
for both self-consistent and relaxed brightness ellipsoid models (same as with our DISC model), so again we only report here the
self-consistent version of the model. The rotational state evolution here is the opposite to that of the DISC model: it starts
as a SAM rotator (rotation and precession periods 80.8 and 8.56~h, respectively), and spins up to become a LAM rotator,
with 2.7 times shorter rotation period and almost unchanged precession period at the end of simulations. Assuming $p=0.1$ albedo,
the physical dimensions are estimated as $324\times 42\times 42$~meters.

\begin{table}
\caption{Confidence intervals for model parameters.}
\label{tab:intervals}
\begin{tabular}{lcc}
\hline
Parameter        & DISC                                                                                                               & CIGAR \\
\hline
$\theta_L$, deg  & $130^{+33}_{-22} \left[\rule{0pt}{0ex}^{+41}_{-22} \rule{0pt}{2.5ex}\right]$ \rule{0pt}{3.5ex}                     & $115.9^{+6.3}_{-1.5} \left[\rule{0pt}{0ex}^{+51}_{-64} \rule{0pt}{2.5ex}\right]$\\
$\varphi_L$, deg & $92.4^{+1.9}_{-8.6} \left[\rule{0pt}{0ex}^{+24}_{-66} \rule{0pt}{2.5ex}\right]$ \rule{0pt}{3.5ex}                  & $112.8^{+3.1}_{-3.1} \left[\rule{0pt}{0ex}^{+68}_{-64} \rule{0pt}{2.5ex}\right]$\\
$\varphi_0$, deg & $357.8^{+1.6}_{-8.4} \left[\rule{0pt}{0ex}^{+37}_{-57} \rule{0pt}{2.5ex}\right]$ \rule{0pt}{3.5ex}                 & $169.6^{+2.0}_{-2.5} \left[\rule{0pt}{0ex}^{+69}_{-97} \rule{0pt}{2.5ex}\right]$\\
$K$              & $7.572^{+0.022}_{-0.080} \left[\rule{0pt}{0ex}^{+5.2}_{-4.8} \rule{0pt}{2.5ex}\right]$ \rule{0pt}{3.5ex}           & $234.6^{+1.6}_{-1.3} \left[\rule{0pt}{0ex}^{+402}_{-208} \rule{0pt}{2.5ex}\right]$\\
$\theta_K$, deg  & $140.33^{+0.44}_{-0.09} \left[\rule{0pt}{0ex}^{+19}_{-61} \rule{0pt}{2.5ex}\right]$ \rule{0pt}{3.5ex}              & $89.753^{+0.003}_{-0.137} \left[\rule{0pt}{0ex}^{+6.1}_{-6.8} \rule{0pt}{2.5ex}\right]$\\
$\varphi_K$, deg & $208.00^{+0.24}_{-0.26} \left[\rule{0pt}{0ex}^{+26}_{-68} \rule{0pt}{2.5ex}\right]$ \rule{0pt}{3.5ex}              & $288.959^{+0.102}_{-0.044} \left[\rule{0pt}{0ex}^{+46}_{-65} \rule{0pt}{2.5ex}\right]$\\
$c$              & $0.1629^{+0.0019}_{-0.0041} \left[\rule{0pt}{0ex}^{+0.165}_{-0.057} \rule{0pt}{2.5ex}\right]$ \rule{0pt}{3.5ex}    & $0.129719^{+0.000040}_{-0.000077} \left[\rule{0pt}{0ex}^{+0.173}_{-0.037} \rule{0pt}{2.5ex}\right]$\\
$b$              & $0.96427^{+0.00024}_{-0.00054} \left[\rule{0pt}{0ex}^{+0.036}_{-0.107} \rule{0pt}{2.5ex}\right]$ \rule{0pt}{3.5ex} & $0.131436^{+0.000056}_{-0.000103} \left[\rule{0pt}{0ex}^{+0.193}_{-0.038} \rule{0pt}{2.5ex}\right]$\\
$E'_0$           & $0.98204^{+0.00038}_{-0.00011} \left[\rule{0pt}{0ex}^{+0.018}_{-0.085} \rule{0pt}{2.5ex}\right]$ \rule{0pt}{3.5ex} & $0.033526^{+0.000001}_{-0.000001} \left[\rule{0pt}{0ex}^{+0.145}_{-0.016} \rule{0pt}{2.5ex}\right]$\\
$L_0$            & $17.050^{+0.011}_{-0.039} \left[\rule{0pt}{0ex}^{+11.2}_{-2.0} \rule{0pt}{2.5ex}\right]$ \rule{0pt}{3.5ex}         & $525.65^{+0.35}_{-0.55} \left[\rule{0pt}{0ex}^{+434}_{-390} \rule{0pt}{2.5ex}\right]$\\
$\psi_0$, deg    & $114.45^{+0.23}_{-0.73} \left[\rule{0pt}{0ex}^{+51}_{-44} \rule{0pt}{2.5ex}\right]$ \rule{0pt}{3.5ex}              & $5.13^{+0.18}_{-0.20} \left[\rule{0pt}{0ex}^{+31}_{-20} \rule{0pt}{2.5ex}\right]$\\
\hline                                                            
\end{tabular}                                                     

\justify
{\it Note.} For each model parameter, we show the
optimal value (from Table~\ref{tab:models}), and two confidence intervals: the constrained one, and the unconstrained one (in square brackets). 
The units are the same as in Table~\ref{tab:models}, except for
angles which were converted to degrees. Instead of the Cartesian normalized torque vector components
$T_{b,c,a}$, here we show optimal values and confidence intervals for torque vector components in the comoving spherical coordinate system: 
$K$ (absolute magnitude), $\theta_K$ (polar angle), and $\varphi_K$ (azimuthal angle).
\end{table}

\begin{figure}
    \includegraphics*[width=\columnwidth]{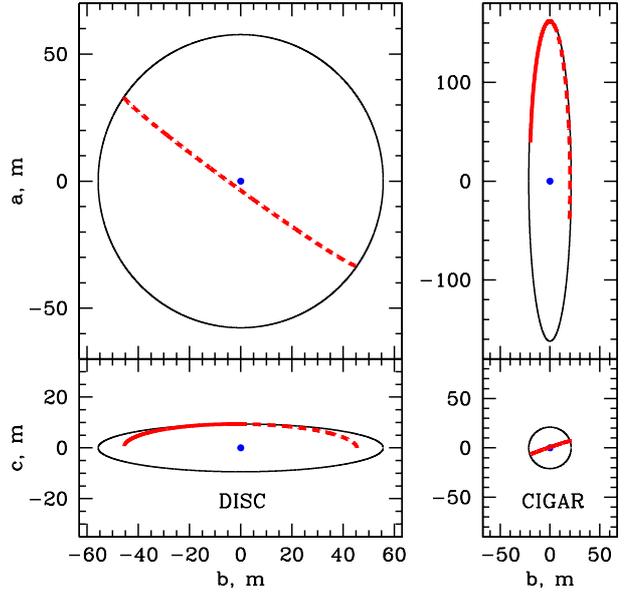}
    \caption{Projection of two models -- DISC (left panel) and CIGAR (right panel) -- onto bOa (top) and bOc (bottom) planes.
    Thin black lines show the extent of the asteroid. Thick red (grey in the printed version of the journal) lines correspond to possible locations of the outgassing point.    
    (The invisible -- behind the asteroid's body -- parts are shown as dashed lines.) 
    The dots show the centre of asteroid. We assumed geometric albedo $p=0.1$.
    }
    \label{fig:torque_map}
\end{figure}

Following \citet{Bartczak2019}, we used our code to estimate the uncertainties in DISC and CIGAR models' parameter determination. Specifically, we computed
weighted root-mean-square deviation (RMSD) of the model curve from the observed one, for models in the vicinity of our best-fitting models.
Only the models which RMSD value was no more than $\mathcal{E}$ above the RMSD value for the best-fitting model were considered to be acceptable.
(Here $\mathcal{E}={\rm RMSD}/\sqrt{N-n}$, where $N$ is the number of data points and $n$ is the number of the model's degrees of freedom.)
In the simplest (and easiest to compute) application of this method, we estimated confidence intervals for model parameters by varying one parameter at a time,
while keeping the rest of parameters fixed at their best-fitting values (from Table~\ref{tab:models}). Our Table~\ref{tab:intervals} lists
these intervals next to the optimal values of the model parameters. As one can see, with the possible exception of the angles $\theta_L$, $\varphi_L$, and
$\varphi_0$, constrained confidence intervals for our model parameters are very small. 

Situation is very different for full (unconstrained)
confidence intervals, which were derived by varying all model parameters simultaneously (values in square brackets in Table~\ref{tab:intervals}).
This difference stems from significant degeneracies between different parameters present in our model.
The unconstrained case computations are dramatically more computationally expensive as they involve exploring the vicinity of a best-fitting model in 11-dimensional model parameter space.
From this analysis, zero torque ($K=0$) models are ruled out for both disc and cigar cases. Torque vector orientation in the comoving coordinte system (angles $\theta_K$ and $\varphi_K$)
is constrained rather poorly, but it is not arbitrary. Same is true for the rest of the angular parameters: $\theta_L$, $\phi_L$, $\varphi_0$, and $\psi_0$.
For DISC model, $b=1$ value (corresponding to an axially symmetric disc) is within the confidence interval. Confidence intervals for parameter $c$ correspond to
the following ranges of the aspect ratios of the models: 1:(3.1 \ldots 9.5) and 1:(3.3 \ldots 10.8), for DISC and CIGAR models, respectively.

Assuming that the torque present in our models is generated by outgassing from one point on the asteroid's surface, we can identify the locus of the plausible
locations of this point as follows.

Applying steady force per unit mass $\mathbfit{f}$ to a point on the asteroid's surface will produce in a general case both constant linear 
acceleration for the whole body, $\mathbfit{f}_r$, and constant tangential (torque) acceleration, $\mathbfit{f}_t$. The linear component is derived by projecting
the vector $\mathbfit{f}$ onto the asteroidal radius-vector at this point, $\mathbfit{r}$; 
the torque component $\mathbfit{f}_t$ is derived by projecting $\mathbfit{f}$
onto the plane perpendicular to the radius-vector. Torque pseudo vector is a cross product of the radius-vector and the tangential component of the force vector, 
$\mathbfit{K}=\mathbfit{r}\boldsymbol\times\mathbfit{f}_t$, and as such is perpendicular to both. As a consequence, the only points on the surface of the ellipsoid 
where a given torque vector can be reproduced are the ones where the radius-vector $\mathbfit{r}$ is perpendicular to $\mathbfit{K}$.
These points lay along the intersection of the plane which is passing through the center of the ellipsoid and is perpendicular to $\mathbfit{K}$.
The intersection line is an ellipse on this plane.
(The full force vector $\mathbfit{f}$ has to lay in the same plane, as its both components -- $\mathbfit{f}_r$ and $\mathbfit{f}_t$ -- lay in that plane.)
Points which are closer to the centre of the ellipsoid would require larger tangential acceleration, points further from the centre
would need smaller tangential acceleration: $f_t=K/r$. (Here $K$ is the modulus of the torque vector.) Assuming that the source of the torque
is the rocket force from outgassing (which is directed inwards), only the points where the angle between the torque vector $\mathbfit{K}$
and the normal to the surface is larger than 90$^{\circ}$ would be physically plausible.

In \autoref{fig:torque_map} we show the locus of physically plausible outgassing points on the surface of the asteroid for 
both models (DISC and CIGAR). As one can see, outgassing can be happening over a wide range of the distances from the asteroid's centre
-- from the central area to the very edge of the object. 

As one can see from Figures~\ref{fig:disc} and \ref{fig:cigar},
torque-driven spin-up of the asteroid is not very obvious in the model light curves. To get a better idea whether the model torque values are physically plausible,
it is instructive to compare our best-fitting models with the data on Solar System comets which had their both linear non-gravitational acceleration
and change of the spin (both caused by the same mechanism -- outgassing) measured. \citet{Rafikov2018a} showed that these comets (there are seven in total)
show a clear correlation between torque $K$ (deduced from the rate of change of the spin) and observed linear non-gravitational acceleration $a_r$:

\begin{equation}
    K=\zeta D a_r.
    \label{eq:Rafikov}
\end{equation}

\noindent (Here $D$ is the characteristic size
of the object -- for a sphere it would be its radius, and $\zeta$ is a small proportionality coefficient which the authors call a "lever arm" parameter.)
The log-average value for $\zeta$ is 0.006, with the full range $0.0007\dots 0.03$. \citet{Rafikov2018a} deduced torque values for Solar System comets
based on some simplifying assumptions, but in our case we can get $K$ values directly from the model. \citet{Micheli2018} showed that the 
linear non-gravitational acceleration of `Oumuamua can be described as $a_r \sim 5\times 10^{-6}$~m~s$^{-2} / R_\odot^2$ 
(here $R_\odot$ is the distance from the Sun in au). Our analysis covers a narrow range of $R_\odot=1.36\dots 1.49$~au. Using the average value for 
$R_\odot$, 1.43~au, we estimate the linear non-gravitational acceleration to be $a_r = 2.45\times 10^{-6}$~m~s$^{-2}$ within the time interval
of interest. Using our model's semi-major axis $a$ in place of $D$, the model's torque value $K$, and `Oumuamua's value of $a_r$ derived above, we can
satisfy \autoref{eq:Rafikov} if we set $\zeta=0.0046$ and $\zeta=0.014$ for DISC and CIGAR models, respectively. This is well within the range of the $\zeta$ values
deduced for Solar System comets, with the DISC value of 0.0046 being close to the log-average $\zeta$ value of $0.006$.
Based on our analysis, our model torque values are consistent with being produced by the same outgassing which presumably drives the
linear non-gravitational acceleration of `Oumuamua.

There is an important caveat in the above analysis: taking our model torque assumptions (torque being fixed in time and space, in the asteroidal coordinate system)
literally, one cannot produce cumulative linear acceleration for the asteroid: as the asteroid spins (with the outgassing point attached to
its surface), the contributions to $a_r$ at different rotation phases would all cancel out. But there are ways to relax our model
assumptions somewhat to circumvent this difficulty. For example, if we make a reasonable assumption that outgassing is the most active when the outgassing
point is facing the Sun, the kinematic model will not change significantly (with our $\mathbfit{K}$ parameter now representing a time-averaged
value of the torque vector), but the linear acceleration can now gradually accumulate, with the
acceleration vector pointing in the right (anti-Sun) direction.

\subsection{Cigar or disc?}
\label{sec:prob}

\begin{figure}
    \begin{center}
    \includegraphics*[width=6cm,keepaspectratio]{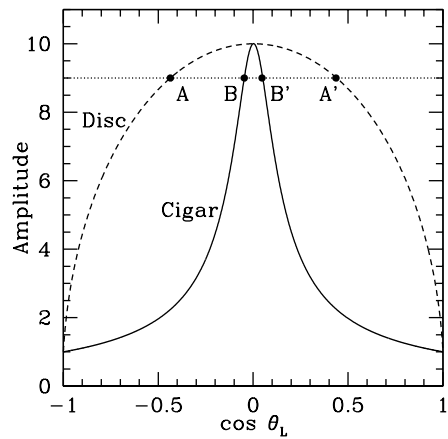}
    \caption{Amplitude of the brightness fluctuations for a 1:10 ratio disc (dashed line) and cigar (solid line), as a function
    of the cosine of the polar angle $\theta_L$ for the angular momentum vector. 
    }
    \label{fig:fluct}
    \end{center}
\end{figure}

The analysis presented in \autoref{sec:torque} demonstrated that the two most promising candidates for a model of `Oumuamua are either a thin disc-shaped or a thin cigar-shaped object subject
to some torque. Unfortunately, it is not possible to differentiate between these two very different cases based solely on the quality of fit of the model light curve to the observed one.

This model degeneracy can be broken by performing a statistical analysis of a different kind, based on the following simple geometric considerations, under the assumption that the
initial angular momentum orientation is random (which is a sensible assumption for an interstellar visitor).
Specifically, to produce large brightness fluctuations (comparable to the asteroid's
largest-to-shortest axes ratio), a cigar-shaped object spinning around its shortest axis would need to have its longest axis repeatedly pointing at the observer with a high accuracy. 
As a consequence, such an object would require a high degree of fine-tuning for its angular momentum vector orientation to produce the desired effect.
The opposite is true for a disc-like
object (at least for the case when it is a LAM rotator): there is a fairly narrow range of the angular momentum vector orientations (when the vector is pointing
towards the observer) when the observer would {\it not} see large brightness fluctuations.

This effect can be easily quantified for the idealistic situation when the object (either a LAM disc or a SAM cigar)
is not tumbling, is not subject to torque, and when we ignore phase effects (by assuming the phase angle is equal to zero). Let us assume that the 
cosine of the angular momentum polar angle $\theta_L$ is equal to zero when the vector is in the plane of the sky (this will result in largest
brightness fluctuations for both disc and cigar). \autoref{fig:fluct} shows how the amplitude of the brightness fluctuations changes as a function 
of $\cos\theta_L$ for 1:10 ratio disc and cigar. (We used the LS ellipsoid
brightness model to compute the brightness; see \autoref{sec:br_ellipsoid}.) For a randomly oriented angular momentum vector $\mathbfit{L}$, 
described by its polar angle $\theta_L$ and azimuthal angle $\varphi_L$,
equal intervals in $\cos\theta_L$ correspond to equal probabilities. From \autoref{fig:fluct} one can see that the disc model is much more likely to produce brightness fluctuations larger than a given
amplitude than the cigar model. For example, amplitudes equal to or larger than nine (horizontal dotted line) will occur in 44~per cent of all disc model cases 
(the length of the interval $A-A'$ divided by two -- the full
range of $\cos\theta_L$), whereas for an equally thin cigar this will be the case in less than 5~per cent of random angular momentum orientations (the length of the interval $B-B'$ divided by two).

\begin{table}
\begin{center}
\caption{Ranked observed brightness minima.}
\label{tab:minima}
\begin{tabular}{lcc}
\hline
Rank & Depth, mag & Feature \\
\hline
1    & 25.715 & D \\
2    & 25.254 & E \\
3    & 25.234 & C \\
4    & 25.212 & A \\
5    & 24.940 & B \\
6    & 24.846 & F \\
7    & 24.834 & L \\
\hline                                                            
\end{tabular}                                                     
\end{center}

{\it Note.} Higher rank corresponds to deeper minimum. "Depth" describes the largest absolute magnitude of the brightness minimum.
Observed light curve features A--L are marked on \autoref{fig:inert}.
\end{table}

This effect should manifest itself to a similar degree in more realistic models, which are tumbling, subject to torque, and have phase effects, though it is more difficult to quantify. We designed
the following statistical analysis pipeline which computes probabilities for our models, by answering the following question: "Given that the initial angular momentum vector orientation
and initial precession angle are random, how likely is it that a given model can produce light curve minima as deep as the observed ones?".

\begin{enumerate}
\item As a starting point, we use one of our best-fitting models -- either DISC or CIGAR (see \autoref{tab:models}). 
\item We relax the LS ellipsoid brightness model, by assigning a given "thickness" to either $c'$ parameter (for disc models)  or both $c'$ and $b'$ (for cigar-shaped models).
\item In our numerical code ran on a GPU, we concurrently generate $256^3$ models which have the same model parameters
(except for the parameters $\theta_L$, $\varphi_L$, and $\varphi_0$) and the same physical scale (parameter $\Delta V$) as our initial model. The three variable parameters -- initial angular momentum
vector orientation angles ($\theta_L$, $\varphi_L$) and initial precession angle $\varphi_0$ -- are sampled with 256 different values each. The sampling is equidistant for azimuthal angles ($\varphi_L$
and $\varphi_0$), and equidistant for the cosine of the polar angle $\theta_L$. As a result, each of the $256^3$ models have equal probability.
\item For each of the $256^3$ models we perform the following steps:
\begin{enumerate}
\item We compute the model light curve in absolute magnitudes, and measure the depth (largest magnitude) of the model brightness minima 
located within the following time intervals:
$t=1.045\dots1.118$, $t=1.978\dots2.185$, $t=3.079\dots3.529$, $t=4.093\dots4.514$, $t=5.234\dots5.355$, and $t=6.181\dots6.278$. (Here $t$ is the number of days since ${\rm MJD}=58050$.)
These time intervals correspond to the observed `Oumuamua light curve intervals which are both wide enough and have enough of observed points to make it possible to
resolve a minimum if it happens to be there.
\item We rank the model minima starting with the deepest (largest absolute magnitude) one. 
\item We set the counter of "good" model minima $N_{\rm min}$ to zero.
\item We compare the depth of the deepest model minimum (model rank \#1) with the depth of the deepest observed minimum (observed rank \#1; see \autoref{tab:minima}).
If the model minimum is deeper (that is, if the absolute magnitude is larger), we increment $N_{\rm min}$
by one.
\item We compare the model rank \#2 minimum to the observed \#2 minimum (\autoref{tab:minima}), and increment $N_{\rm min}$ by one if the model minimum is deeper.
\item We repeat the previous step for ranks \#$3\dots7$.
\item A the end, each of the $256^3$ equally likely models will have a value of $N_{\rm min}\in[0,7]$. If $N_{\rm min}=0$ then none of the model minima were as deep as the corresponding rank observed minima.
If $N_{\rm min}=7$ then all of the model minima were deeper than the corresponding rank observed minima.
\end{enumerate}
\item By counting the number of models where $N_{\rm min}=7$ and dividing the number by the total number of models ($256^3$) we can estimate how likely the initial model is.
\end{enumerate}

\begin{table}
\begin{center}
\caption{Probabilities for different models.}
\label{tab:prob}
\begin{tabular}{lccc}
\hline
Model    & Thickness & $\langle N_{\rm min}\rangle$  & Probability \\
\hline
DISC     & 0.139     & 5.89         & 0.50       \\
         & 0.10      & 6.84         & 0.91       \\
         & 0.05      & 6.86         & 0.92       \\
         & 0.01      & 6.87         & 0.92       \\
\hline
CIGAR    & 0.10      & 1.37         & 0.16       \\
         & 0.05      & 1.55         & 0.20       \\
         & 0.01      & 1.60         & 0.21       \\
\hline                                                            
\end{tabular}                                                     
\end{center}

{\it Note.} Thickness parameter describes $c'$ in the disc model, and both $c'$ and $b'$ in the cigar model. $\langle N_{\rm min}\rangle$ is the average number of ranked model minima which are 
deeper than the seven observed ranked minima (the allowed range is $0\dots 7$). 
\end{table}

We performed the above analysis for our two best-fitting torque models -- DISC and CIGAR -- for a few different thickness values. The results are summarized
in \autoref{tab:prob}. The most striking result here is that regardless of how thin the model is (from the plausible value of $0.10$ down to the extreme value of 0.01), disc models are very likely
(in fact almost guaranteed, with $\sim 91$~per cent probability) to produce brightness minima as deep as observed `Oumuamua's minima. Cigar models, on the other hand, are very unlikely to
reproduce the deep observed minima, with the probability of only 16~per cent for the plausible thickness of 0.10 (which grows only slightly -- to 21~per cent -- for the implausible thickness of 0.01).
Based on this statistical analysis (which is independent from the $\chi^2$ goodness-of-fit analysis we performed in the previous section), the expected thickness of disc models is $\sim 0.14$
(when the probability of the DISC model is around 50~per cent; the first line in \autoref{tab:prob}), which is slightly smaller than the value derived by means of $\chi^2$ fitting ($c=0.16$, from \autoref{tab:models}).

\begin{figure*}
 \begin{multicols}{2}
 \includegraphics*[width=\columnwidth]{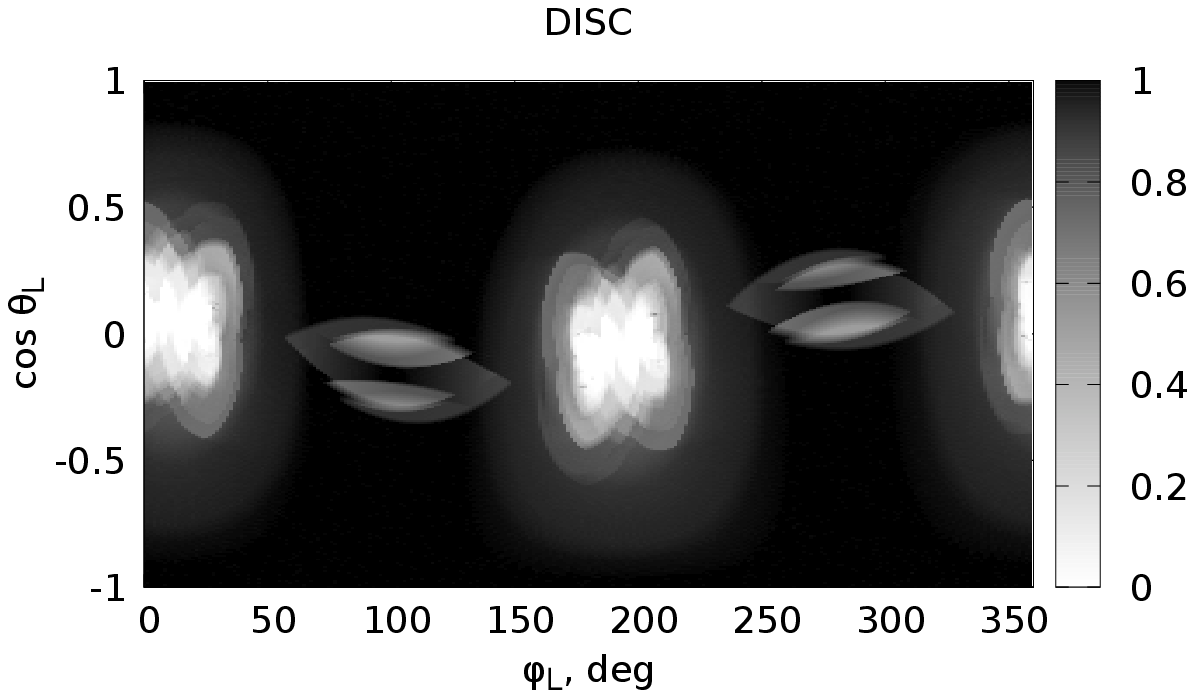}\par
 \includegraphics*[width=\columnwidth]{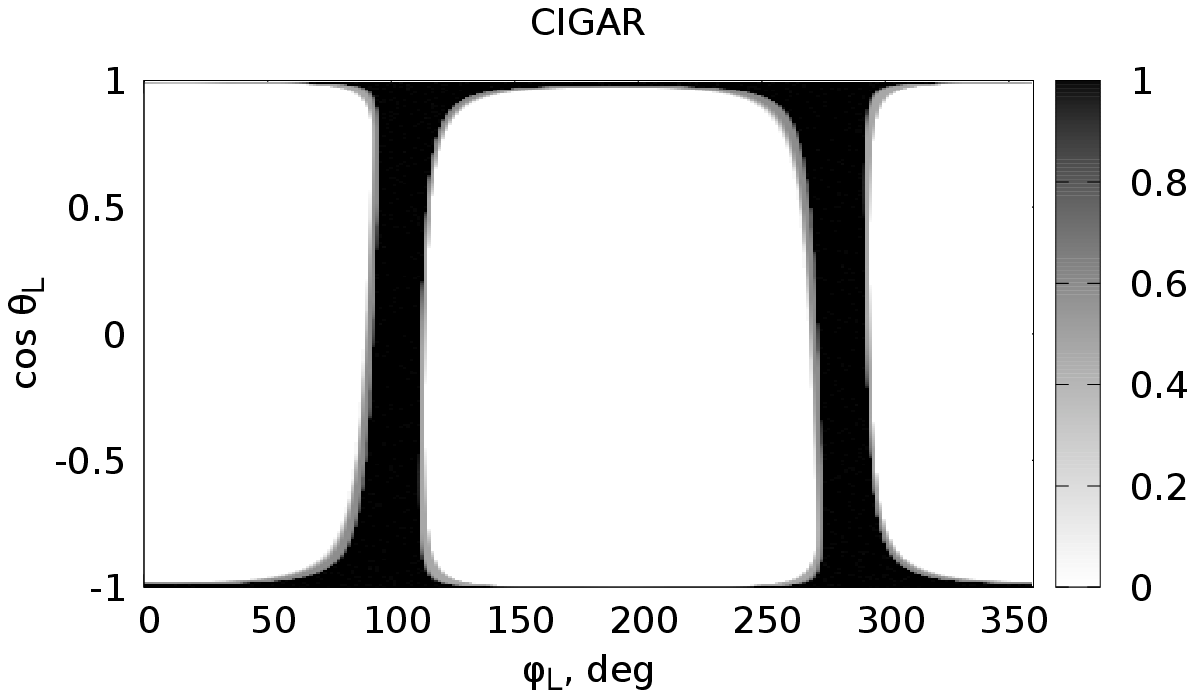}\par
 \end{multicols}
    \caption{Probability maps for our best-fitting models (DISC on the left and CIGAR on the right). We set the thickness parameter to 0.1 for both models.
    Polar ($\varphi_L$) and azimuthal ($\theta_L$) angles
    describe the initial orientation of the angular momentum vector. Assuming that this vector is oriented randomly, each pixel in these maps is equally likely.
    Each pixel of the map presents the model probability averaged over all values of the initial precession angle
    $\varphi_0$. Black corresponds to 100~per cent probability, white corresponds to 0~per cent probability.}
    \label{fig:map}
\end{figure*}

In addition to producing a single probability value for each model, it is instructive to analyse detailed probability maps for the initial angular momentum vector orientation
(\autoref{fig:map}). As expected, the disc model is very likely for almost any orientation
    of the angular momentum vector, whereas the cigar model requires a high degree of fine-tuning in terms of the angular momentum vector orientation.

Based on the statistical analysis presented in this section, `Oumuamua is most likely a disc-shaped object, though the cigar shape cannot be ruled out. 
Interestingly, \citet{Sekanina2019a} reached a similar conclusion -- that a disc shape is much more likely than a cigar one -- based on a completely different argument
(non-detection of `Oumuamua by the {\it Spitzer Space Telescope}).

\subsection{Alternative models}
\label{sec:alternative}

In this section we consider two additional auxiliary models for `Oumuamua.

The first auxiliary model is identical to our fiducial model (LS ellipsoid brightness model + constant torque; \autoref{sec:torque})
in all aspects except for numerical values of some parameters -- the thickness (parameter $c$),
which now has the initial range (soft limits) between $10^{-4}$ and $10^{-2}$ (the range was $[0.01,1]$ in the fiducial model), and the length of  the intermediate 
semi-axis $b$ (new soft limits: $[0.3,1]$). 
These numerical runs were an attempt to use our model's framework to explore the idea of \citet{Bialy2018} that 
`Oumuamua is a solar sail -- an extremely thin object, with very low surface density ($\sim 0.1$~g~cm$^{-2}$). 

As we already demonstrated that there is no need to use relaxed brightness ellipsoid with the models with constant torque to reproduce well `Oumuamua's light curve (\autoref{sec:torque}),
we used the same self-consistent brightness ellipsoid in the "solar sail" simulations. That means the total
number of free parameters was 11 (8 basic tumbling model parameters plus 3 torque component parameters). We carried out our standard set of numerical runs (\autoref{sec:runs}) for the new model.

\begin{figure}
    \includegraphics*[width=\columnwidth]{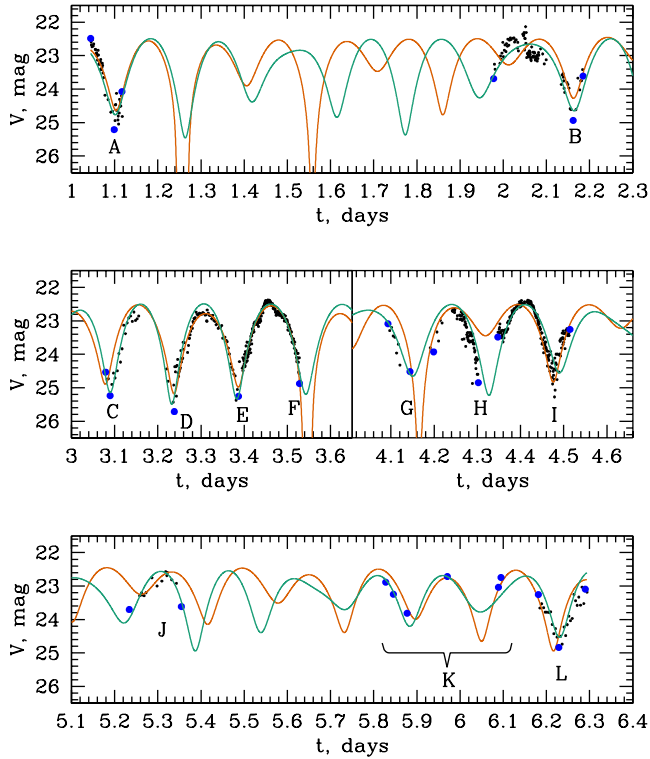}
    \caption{Our best-fitting alternative models: SAIL (orange line; dotted line in the printed version of the journal) and BALL (green line; solid line in the printed version of the journal).
    See the caption of \autoref{fig:inert} for details.
    }
    \label{fig:alt}
\end{figure}

Our main finding here is that the quality of the light curve fit for the "solar sail" models is noticeably worse than for our best-fitting fiducial models (DISC and CIGAR). As one can see from
\autoref{fig:alt}, our best-fitting model (SAIL) struggles to reproduce the brightness maximum features J and the one around $t=2.0$,
the depth of the minimum H, and has timing issues with the minima C and L. The model does require some torque (\autoref{tab:models}). It is interesting that the model is degenerate
in the sense that the light curve is essentially unchanged starting from $c\sim 0.001$ down to the smallest value we tested ($10^{-5}$; that is the value we use in \autoref{fig:alt}
and \autoref{tab:models}). As this is a self-consistent brightness ellipsoid model, the degeneracy is present for both kinematic and brightness ellipsoids. As a result,
our SAIL model is consistent with the extremely low surface density requirement of the solar sail hypothesis of \citet{Bialy2018}. 

It is easy to see why the model becomes degenerate in the $c\rightarrow 0$ and $c'\rightarrow 0$ limits. The kinematic part of the model becomes degenerate because as $c$ is approaching zero,
the three diagonal components of the inertia tensor converge to constant values (see \autoref{eq:Iabc}): $I_a=1$, $I_b=b^{-2}$, $I_c=(1+b^2)b^{-2}$. This in turn ensures that the kinematic ODEs
(\autoref{eq:euler2}) no longer depend on $c$ (become degenerate). Even stronger effect is observed in the brightness model part. In the limit of $c'\rightarrow 0$ our triaxial
ellipsoid model degenerates into an extremely thin flat object. One can easily show that neither different shapes nor variable albedo can affect 
the light curve from such an object, as each individual element of the object's surface would see exactly the same phase angle as other elements on the same side of the sail. 
Changing the shape of the object by moving the elements around (in the sail's plane) will not affect the integrated brightness at any given orientation of the sail. Variable albedo
is also incapable of changing the light curve shape, as at each orientation of the sail all elements (with different albedo) would contribute to the integrated brightness in the
same proportion. The only factors which still can affect the shape of the light curve for a solar sail (and hence can be potentially used to further improve the quality of fit
between the model and observed light curves) are (a) a different torque model, and (b) a different light scattering model. A more fundamental change would be to assume
that the thin sail is not perfectly flat (it has some curvature or ripples). Modeling these effects would go beyond the scope of this paper.

Assuming a fairly high albedo of 0.5 (which would be appropriate for
a solar sail), the size of the SAIL disc is $64\times 51$~m (full diameters). The torque "lever arm" parameter $\zeta$ is $0.017$, implying that less than 2~per cent of the linear 
non-gravitational force experienced by `Oumuamua needs to be converted to torque. 

How can radiation pressure produce the required torque? We speculate that making the shape
and/or mass distribution asymmetric may not do the trick, as once the sail makes half a full rotation, the opposite direction torque would be exerted, canceling out
the original torque. Variable albedo seems to be a more promising agent. Let us assume `Oumuamua is a flat disc-shaped sail of radius $R$ with one of the sides consisting of a darker
(albedo $p_1$) and brighter (albedo $p_2$) halves. Radiation pressure can be computed as $P=(1+p)C$, where $p$ is the albedo and $C$ is a constant when the distance from the Sun
is fixed \citep{Bialy2018}. An element with the surface area $\mathrm{d}S$ will experience force $\mathrm{d}f=P\,\mathrm{d}S$. Integrating the product $r\,\mathrm{d}f$ over the darker half of the disc 
($r$ being the distance of the element from the disc centre)
gives us the torque applied to that side, $K_1=(1+p_1)C\pi R^3/3$; doing the same for the brighter side gives us the second torque component, $K_2=(1+p_2)C\pi R^3/3$. 
The global torque is $K=K_2-K_1=(p_2-p_1)C\pi R^3/3$, while the global linear acceleration due to solar radiation is $a_r=(1+\langle p\rangle)C\pi R^2$.
(Here $\langle p\rangle=(p_1+p_2)/2$ is the average albedo of the side.) Substituting the above expressions for $K$ and $a_r$ into \autoref{eq:Rafikov} (and setting $D=R$), we derive the following
expression for the difference between the two albedos which will generate the required torque: $p_2-p_1=3(1+\langle p\rangle)\zeta$. Assuming
$\langle p\rangle=0.5$, we can produce the required torque ($\zeta=0.017$) if the albedos differ by a fairly small amount: $p_2-p_1\sim 0.08$.

Our second auxiliary model uses a completely different brightness model. Specifically, it explores an alternative (non-geometric) explanation for the large brightness variations of `Oumuamua, where
the asteroid is assumed to be roughly spherical in shape but with large albedo variations across its surface. We use the simplest possible implementation of this idea -- "black-and-white ball" 
brightness model (\autoref{sec:BW}), with only three free parameters -- polar coordinates $\theta_h$, $\varphi_h$ of the dark spot on the surface of the asteroid (in the comoving coordinate system
$bca$), and the dark/bright sides albedo ratio $\kappa$. Both dark and bright sides are equal in size (both are hemispheres) for simplicity. Also, we ignore phase
effects (phase angle is assumed to be zero).

Our primary motivation to explore non-geometric explanations for the large brightness variations of `Oumuamua was the fact that this approach presents a completely different coupling
between the spinning/tumbling motion and the brightness variations. Specifically, in our "black-and-white ball" model there is one maximum and one minimum in the light curve
per full rotation of the body. This is in contrast with the geometric (ellipsoid) picture, where one has two maxima and two minima per rotation. Our hope was that this very different coupling
might remove the need for torque when trying to fit the observed light curve of `Oumuamua.

Our analysis showed this not to be the case. Running our full suite of simulations for an inertial tumbling "black-and-white ball" failed to produce light curves which were a noticeably better fit
to `Oumuamua's observed light curve than with our inertial LS brightness ellipsoid runs (\autoref{sec:inertial}). (For the model parameter $\kappa$ we used logarithmic scaling and soft
limits $[0.01,0.1]$; the kinematic ellipsoid's thickness $c$ had soft limits $[0.3,1]$.) Crucially, similarly to the inertial brightness ellipsoid case, the best-fitting zero torque 
"black-and-white ball" models had serious issues matching the timings of the observed light curve minima.

Adding constant torque to the above model rectified the situation (the same way it helped in the LS brightness ellipsoid simulations). 
We ran a full suite of numerical simulations for a "black-and-white ball"
with torque (14 free model parameters in total: 8 basic tumbling model parameters plus 3 "black-and-white ball" brightness model parameters plus 3 torque components), and show our best-fitting
model BALL in \autoref{tab:models} and \autoref{fig:alt}. Now the timings of the model minima match well those of the observed minima, and overall quality of fit is decent (in fact 
better than for our models DISC and CIGAR). The model BALL starts as a SAM rotator with the rotation period of 23.9~h and the precession period of 4.5~h, and ends up as a LAM rotator
with $P_\psi=20.1$~h and $P_\varphi=3.9$~h after 5~days. The parameter $\kappa$ (dark/bright sides albedo ratio) is 0.03. The kinematic ellipsoid is not exactly a "ball", 
but with the shape parameters $c=0.52$ and $b=0.93$ its geometry is much less extreme than in our models DISC and CIGAR. 
Polar angle $\theta_h$ is equal to 162$^\circ$, meaning that the dark hemisphere is fairly close to the  "southern polar region" of the "ball".

\section{Discussion}
\label{sec:discussion}

Prior attempts to interpret `Oumuamua's light curve \citep{Fraser2018,Belton2018,Drahus2018} were based on searching dominant frequencies and interpreting them as a linear combination
of two frequencies  -- precessional and rotational. This approach usually works very well for Solar System asteroids and comets, but its fundamental assumption is that torque is zero. If that is
not the case, the dominant frequencies found in light curves can no longer be used to find the rotational state of the asteroid: at best they might correspond to real frequencies present in
some segments of the data which are particularly well sampled, at worst they are purely fake, reflecting the patchiness of the data. \citet{Samarasinha2015} provide one such example, when adding
noise to the perfect model data and making it patchy produced fake dominant frequencies.

Our research represents the first attempt to fit `Oumuamua's light curve using a physical model. (Recently published research by \citealt{Seligman2019} did use a physical model
with torque to explain `Oumuamua's light curve, but they did not carry out multi-dimensional model fitting, so their results are only suggestive; 
the computational tasks are completely incomparable: where we had to compute hundreds of millions of physical models, they only computed a few.)
The fundamental advantage of such an approach is that torque can be
modeled directly. In addition, other aspects affecting the light curve (variable phase angle, different shapes, spatially variable albedo etc.) can also be directly modeled, which removes a lot
of guesswork from interpreting light curves.

We started this project fully expecting that given how limited, noisy, and patchy `Oumuamua's light curve is, we would be finding a large number of very different inertial models which would
all provide a comparable quality and reasonable description of the data. The first big surprise was when we realized that no inertial model we tried (LS ellipsoid, "black-and-white ball", "solar sail") 
could match the timings of the most conspicuous features of the observed light curve -- the multiple deep and narrow minima. The simplest non-inertial extension of the model we tried
(steady torque fixed in the comoving coordinate system) was sufficient to rectify this situation for all of our brightness models. In all likelihood our torque model, with only 
three free parameters, is an oversimplification, as any realistic mechanism producing torque would be significantly more complicated (time variable, not firmly attached to the surface
of the object etc.). The important point here is that any torque prescription, even as simple as the one we used, should be able to fix the minima matching issues which plagued all our
inertial models. 

We consider the finding that some torque is needed to model well the light curve of `Oumuamua to be our main result. It is quite remarkable that the torque
required is in line with the results for the Solar System comets for which both linear non-gravitation acceleration and change of the spin (both effects driven by outgassing)
were measured \citep{Rafikov2018a}. This could be viewed as an important evidence supporting the comet hypothesis for `Oumuamua.
We should caution though that this does not prove the non-gravitational acceleration of `Oumuamua and its torque are driven by outgassing. Other mechanisms where a force
is applied to the asteroid's surface can have a comparable relation between the torque and the linear acceleration. For example, solar radiation can drive primarily the linear acceleration
of a thin object ("solar sail"), but can also generate some torque (for example, if albedo varies across one or both sides of the sail; see \autoref{sec:alternative}).

As a side note, we suggest here one mechanism which by design will only produce linear non-gravitational acceleration (or rather an appearance of such), with zero torque: 
if `Oumuamua happens to be made of some sort of exotic matter for which the gravity law deviates slightly from the canonical form. Indeed, if the gravity constant $G$
for `Oumuamua were only $0.0008$ of fractional units smaller than the standard value, this would completely reproduce the effect discovered by \citet{Micheli2018}:
the appearance of an additional force which is radially directed away from the Sun, scales as $r^{-2}$ and has the right magnitude. As this is not a real force, there would be zero torque by design.

Both the discovery of the non-gravitational acceleration by \citet{Micheli2018} and our current findings strongly suggest that `Oumuamua should have experienced a fairly strong torque.
But recently \citet{Rafikov2018b} claimed that `Oumuamua experienced negligible torque, and hence cannot be a comet. We would like to point out an internal inconsistency in the
argument of \citet{Rafikov2018b}. On one hand it is true, as the author claimed, that given that the periodogram analysis of `Oumuamua's light curve (spanning 30 days) carried out by
\citet{Belton2018} revealed the presence of a dominant period of ($8.67\pm0.34$)~h, one possible explanation can be the hypothesis that the frequency is physical
(corresponding to either precessional, rotational periods, or some combination of the two) and that the torque is so weak that the dominant frequency did not change by more
than the quoted uncertainty of 0.34~h over the 30 days. On the other hand, once one assumes the torque is strong enough to affect the spin of `Oumuamua, one can no longer interpret 
dominant frequencies recovered from the light curve as physical. Non-negligible torque would smear out the physical frequencies in the periodogram, leaving instead artefacts of the patchiness of the data, or perhaps a dominant frequency present in the most sampled segment(s) of the light curve. The latter may very well be the case here, as the DISC model minima D and H 
are separated by three times 8.54~h, and minima E and I
are separated by three times 8.78~h (see \autoref{fig:disc}; all four minima are among the best sampled in the light curve). The average of these two periods is 8.66~h (almost exactly
the dominant period of 8.67~h detected by \citealt{Belton2018}), and the deviations from the average are 0.12~h -- well within the uncertainty of 0.34~h of the detected period. This period
could conceivably show up in a periodogram for the model's light curve, despite the fact that the model lacks a well defined period due to the effects of steady torque.

Staying within the realm of conventional explanations for `Oumuamua (asteroid versus comet), both the presence and magnitude of torque evidenced by the current research would appear to tilt 
the scales towards the cometary nature of the object. One has to emphasize though that the lack of any direct signs of outgassing for `Oumuamua is highly troubling. Trying to reconcile
the cometary hypothesis with the lack of outgassing detections, \citet{Micheli2018} had to assume a rather extreme composition of the object in terms of the CN to H$_2$O ratio and the dust
properties, leaving the H$_2$O and CO as the most likely drivers of the non-gravitational acceleration of the asteroid. The non-detection of CO outgassing using {\it Spitzer Space Telescope}
\citep{Trilling2018} and the argument of
\citet{Sekanina2019} that H$_2$O has much lower abundance than what is needed to drive the non-gravitational
acceleration of `Oumuamua make the cometary explanation even more problematic. If `Oumuamua is a comet in some sense, it must be a very exotic one, with its properties (chemical composition and
geometry) being nothing like properties of Solar System comets.

This makes other ("exotic") explanations for `Oumuamua's nature quite competitive. Even though our model SAIL, designed to mimic the solar sail hypothesis of \citealt{Bialy2018}, does not provide
as good fit to `Oumuamua's light curve as our more conventional models, DISC and CIGAR, relaxing some of our model assumptions (e.g. changing the light scattering law,
or assuming that the thin sail has a curvature or ripples) could potentially make it a viable option.
Importantly, our model is degenerate, allowing the thickness of the object
to be arbitrarily small -- even as small as the solar sail requirement, $~\sim 0.5$~mm. The model does require some torque to match the timings of the asteroid's brightness minima
reasonably well, but as we argued earlier solar radiation can generate torque if the albedo varies across the surface of the sail.

Another (semi)-exotic explanation for `Oumuamua we considered -- a "black-and-white ball" -- was a failure in the sense that it did not remove the need for torque. The model
has a rather extreme bright-do-dark sides albedo ratio of 32. Given that `Oumuamua did not exhibit obvious color variations (with the possible exception of a "red spot", as noted by
\citealt{Fraser2018}), and that for Solar System minor bodies shape is the main driver of large brightness variations, this hypothesis should be treated as an interesting but unlikely alternative
explanation for the asteroid's nature. 

As our second main result, in this paper we presented the evidence that by far the most likely shape for `Oumuamua is a disc (or slab, or pancake). 
Making a reasonable assumption that `Oumuamua's angular momentum vector had no preferred direction, the requirement for the model to produce light curve minima as deep as the observed ones
sets the likelihood of the cigar shape, popular in the literature, at
only 16~per cent.
A thin disc, on the other hand, is very likely to produce brightness minima of the required depth. 
Disc-shaped and cigar-shaped
objects produce very similar-looking light curves (compare \autoref{fig:disc} and \autoref{fig:cigar}). It takes a statistical analysis of a different kind (presented in \autoref{sec:prob}) 
to break this model degeneracy. This finding may have interesting implications for future discussions 
about the nature of the asteroid. In particular, recent research providing explanations for `Oumuamua's cigar shape (e.g. \citealt{Katz2018,Vavilov2019,Sugiura2019}) may need to be revisited.

Combining our physical model fitting of `Oumuamua's light curve with our statistical analysis of the model probability based on the depth of the light curve minima points to a tumbling thin disc
experiencing some torque as the most likely model for the asteroid. The disc diameter is $\sim 110$~m (assuming geometric albedo $p=0.1$), and it is very close to being axially symmetric.
The model is self-consistent (the same ellipsoidal shape explains both the kinematics and the brightness variations).
The disc thickness is estimated at 19~m (from the light curve fitting) or 16~m (from the probabilistic minima depth analysis; see \autoref{tab:prob}).
It requires a moderate amount of torque over the 5 days covered by this analysis, consistent with the amount of torque experienced by Solar System comets. The remaining deviations of the
model light curve from the observed one suggest that the shape of the object is not exactly ellipsoidal and/or there are some albedo variations across its surface.

Our analysis only covered a very short time interval (5 days). One important question is: what is the longer term impact of torque in our models? Will the asteroid 
spin up in a fairly short time to the point that it breaks apart? To start with, our model assumption of steady torque fixed in the asteroidal coordinate system is a significant
oversimplification. At the very least, it should go down as $r^{-2}$ as the object moves away from the Sun. Also, as we discussed earlier (\autoref{sec:torque}), real torque has
to vary with the rotation phase (e.g., by becoming stronger when the outgassing point is heated by the Sun), otherwise the model will not produce the linear acceleration term.
As a worst case scenario, we ran our models DISC and CIGAR for 25 more days (bringing the total evolution time to 30 days), maintaining the same fixed values of the torque
pseudo vector. During this time, the light curves for both models remained fairly regular, extending the trend from the first 5 days (see \autoref{fig:disc} and \autoref{fig:cigar}).
The effective rotation period (interval between alternate minima) for the DISC model changes from 9.6~h to 2.3~h after 30~d. For the CIGAR model the change is much
less steep (from 9.4~h to 8.4~h), which suggests that for this model the torque primarily impacts the direction of the angular momentum vector. Even for the most affected model
(DISC), the rotation period after 30 days (2.3~h) is still short of what is needed to break up the asteroid ($<1$~h; \citealt{Rafikov2018b}). Once one takes into account
the $r^{-2}$ dependence of the torque on the distance from the Sun, the spin up due to torque will be even more moderate.

Our model cannot tell us what was happening before the 5-day interval we simulated. The asteroid was closer to the Sun, so presumably the torque was stronger. It is very likely
that as we move backward in time, if we simply assume the torque direction in the asteroidal coordinate system is fixed, and its magnitude grows as $r^{-2}$, our model would quickly become
unphysical. One way out of this is to assume that over longer time intervals our assumptions that torque is constant and the outgassing point is fixed in the comoving coordinate system can no longer be valid even 
in an approximate sense. A more realistic picture would have multiple outgassing points happening primarily in the Sun-lit parts of the asteroid. The outgassing model of \citet{Seligman2019},
where the outgassing point is not fixed in the asteroidal coordinate system but instead tracks the subsolar point,
may be more appropriate for longer time interval simulations.

\section{Conclusions and Future Work}
\label{sec:conclusions}

We presented the first attempt to fit the light curve of the interstellar asteroid `Oumuamua using a physical model, which consists of the kinematic part (tumbling asteroid subject to
constant torque) and the brightness model part (either Lommel-Seeliger triaxial ellipsoid or "black-and-white ball"). We performed exhaustive, Monte-Carlo style, multi-dimensional optimization 
of the models using our numerical, GPU-based code, developed specifically for this project. We spent approximately one GPU-year for this project, using NVIDIA P100 GPUs.

Here are our main findings.

\begin{enumerate}

\item Some torque is needed to explain the exact timings of the deep light curve minima of `Oumuamua. This is true for all brightness models we tried (LS ellipsoid -- 
including the special case of a "solar sail" -- and "black-and-white ball").

\item The amplitude of the torque required by our best-fitting models is consistent with the torque measured for Solar System comets which spin and radial acceleration was 
affected by outgassing. 

\item Our analysis produced two different best-fitting ellipsoidal models for `Oumuamua: either a thin disc or a thin cigar.
 Both models are very close to being axially symmetric, and are self-consistent
(brightness ellipsoid is identical to the kinematic ellipsoid).

\item Assuming random orientation of the asteroid's angular momentum vector, we computed the probability that our best-fitting models can produce light curve minima as deep as the observed ones.
This analysis demonstrated that the disc shape (probability 91~per cent) is much more likely than the cigar shape (probability 16~per cent). 

\item Our best overall model for `Oumuamua is a thin disc ($115\times 111\times 19$~m assuming geometric albedo $p=0.1$) which is initially a LAM rotator with the rotation
and precession periods of 51.8~h and 10.8~h, respectively. After five days it evolves into a SAM rotator with the rotation
period of 32.3~h (the precession period remains essentially unchanged). The "lever arm" parameter $\zeta$ (the measure of the torque strength in relation to the non-gravitational linear
acceleration of the asteroid) for this model is 0.0046, which is close to the log-average value of 0.006 for Solar System comets.

\item Though we consider the two alternative models we tried ("solar sail" and "black-and-white ball"; both needed some torque) less likely, we believe they are viable.

Our current research has definitely not exhausted the field of physical modeling of `Oumuamua. The asteroid's light curve appears to be rich enough (with multiple sharp features)
to sustain even more advanced physical modeling. In particular, attempts can be made to carry out a full light curve inversion (like in \citealt{Kaasalainen2001b}), 
to try to recover the true shape (with no assumptions of symmetry and convexity) of the asteroid. One could also try to model both the variable shape (e.g. as a triaxial ellipsoid) and albedo
variations across the surface, or try different torque prescriptions (e.g. the one used by \citealt{Seligman2019}).
Finally, more advanced solar sail models (with some curvature and variable albedo) could be developed, with the hope that they can both explain the observed light curve and have self-consistent torque
and linear non-gravitational acceleration (both driven by solar radiation).

\end{enumerate}

\section*{Acknowledgements}

This research was enabled in part by support provided by SHARCNET (www.sharcnet.ca) and Compute Canada (www.computecanada.ca). The computations were carried out on clusters Graham and Cedar,
operated by Compute Canada. The author would like to thank Michal Drahus, Wesley Fraser and Petr Pravec for providing access to light curve data for 'Oumuamua and 2002 TD$_{60}$, Roman Rafikov, Doug Welch, and Zdenek Sekanina for constructive criticism of the manuscript, and Mark Hahn for proofreading the text of the paper. He would also like to thank the paper's referee, Przemyslaw Bartczak, for reviewing the manuscript and very helpful suggestions which improved the quality of the paper.




\bibliographystyle{mnras}
\bibliography{ms} 




\appendix


\bsp	
\label{lastpage}
\end{document}